\documentclass[12pt,preprint]{elsarticle}

\usepackage[utf8]{inputenc}
\usepackage[T1]{fontenc}
\usepackage{amsmath,amssymb,amsfonts,bm}
\usepackage{graphicx}
\usepackage{booktabs}
\usepackage{algorithm}
\usepackage{algpseudocode}
\usepackage{subcaption}
\usepackage[hidelinks]{hyperref}
\usepackage[capitalize,noabbrev]{cleveref}
\usepackage{xcolor}
\usepackage{lineno}

\journal{Journal of Computational Physics}

\begin{document}

\begin{frontmatter}

\title{An Implicit Time-Domain Harmonic Balance Method for Radio-Frequency Capacitively Coupled Plasma Simulations}

\author[a]{Yuze ZHU}
\author[a]{Yufeng WEI}
\author[a]{Yue ZHANG}
\author[a,b,c]{Kun XU\corref{cor1}} 
\cortext[cor1]{Corresponding author. Email: makxu@ust.hk}

\affiliation[a]{
   organization= {Department of Mathematics, Hong Kong University of Science and Technology},
   city= {Clear Water Bay, Kowloon, Hong Kong},
   country= {China},
}

\affiliation[b]{
   organization= {Department of Mechanical and Aerospace Engineering, Hong Kong University of Science and Technology},
   city= {Clear Water Bay, Kowloon, Hong Kong},
   country= {China},
}

\affiliation[c]{
   organization= {Shenzhen Research Institute, Hong Kong University of Science and Technology},
   city= {Shenzhen},
   country= {China},
}

\begin{abstract}
Fast and accurate fluid simulation of radio-frequency capacitively coupled plasmas (RF CCPs) is of great importance for the iterative design and parameter optimization of modern plasma reactors. This study presents the first successful extension of the time-domain harmonic balance (HB) method to a fully coupled drift-diffusion-Poisson system with complete electron-energy transport for RF plasma simulations. To resolve the severe numerical stiffness arising from highly nonlinear energy-dependent kinetics and dense phase-coupling, a highly efficient spatiotemporal operator-splitting strategy is employed. By sequentially executing a spatial implicit relaxation and a cell-local temporal inversion, this strategy entirely avoids the memory-intensive assembly of global Jacobians while preserving robust numerical stability. The proposed method is rigorously validated against a standard parallel-plate argon CCP benchmark. Evaluated across all discrete temporal collocation points, the HB solution demonstrates that retaining eight harmonics ($N_H=8$) perfectly resolves both the quasi-steady bulk plasma and the highly nonlinear transient sheath dynamics, yielding macroscopic relative errors strictly below $0.3\%$ compared to conventional dual-time stepping (DTS) solutions. Beyond its high physical fidelity, the time-domain HB method completely bypasses the prohibitive physical transients required by conventional time-marching methods. Evaluated on a purely sequential single-core execution, the HB method delivers a greater than 10-fold speedup over fully converged DTS baselines and remains over 5 times faster than the coarsest time-marching configurations. These results establish the time-domain HB framework as a physically rigorous, memory-efficient, and highly accelerated paradigm for practical RF plasma simulations.
\end{abstract}


\end{frontmatter}

\section{Introduction}
\label{sec:introduction}

Radio-frequency capacitively coupled plasmas (RF CCPs) are widely used in plasma-assisted etching, deposition, surface activation, and semiconductor manufacturing ~\cite{lieberman1994principles, chen2003lecture, chabert2011physics, oehrlein2018foundations}. Their practical importance has motivated decades of work on kinetic~\cite{birdsall1991particle,surendra1991particle}, hybrid~\cite{kushner2009hybrid, matthews1994current}, and fluid plasma simulation~\cite{misium1989macroscopic, boeuf1987numerical}. Among these approaches, fluid models remain attractive for reactor-scale calculations because they reduce the kinetic degrees of freedom to a comparatively small set of moments, thereby providing the computational efficiency required for routine iterative design and optimization. While the plasma dynamics are driven by periodic excitations with fixed frequencies, the numerical integration must simultaneously capture complex multiphysics and chemical processes, such as electron drift and diffusion, dielectric relaxation, sheath formation, volumetric reaction kinetics, and electron-energy transport. Consequently, conventional time-marching calculations remain computationally demanding: the time step is tightly restricted by the smallest relevant physical or numerical scales, and the computation must be advanced over many RF cycles until the initial transient decays~\cite{godyak1990dynamic, sharma2013investigation}.

The immense computational cost of reaching the periodic steady state has motivated various acceleration strategies \cite{hagelaar2000modeling,arslanbekov2021implicit,gomez2023jacobian}. Fully implicit and semi-implicit drift-diffusion-Poisson solvers can significantly relax the severe restrictions imposed by dielectric-relaxation and explicit transport stability limits. However, taking large time steps yields strongly coupled nonlinear algebraic systems, whose poorly conditioned Jacobians demand carefully designed scaling, linearization, and preconditioning techniques. Alternatively, high-order finite-volume or finite-element spatial discretizations can achieve a target accuracy with significantly fewer spatial degrees of freedom \cite{davoudabadi2009accuracy}. More recently, reduced-order and machine-learning models have been deployed to accelerate repeated parameter queries; yet, because these data-driven approaches rely on high-fidelity time-domain simulations for training, they inherently inherit the prohibitive upfront cost of generating the database \cite{trieschmann2023machine}. Another classical approach to bypass the transient phase is the Newton-Raphson cycle-acceleration method, which formulates a root-finding problem for the periodic state by integrating sensitivity equations over a full RF cycle \cite{lymberopoulos1993fluid}. While this scheme dramatically reduces the number of simulated cycles for species with slow response times, such as metastables, evaluating the full-cycle Jacobian is generally too memory- and CPU-intensive to apply to all degrees of freedom in multidimensional fluid models. Despite these remarkable advancements, completely eliminating the computational waste of time-marching through a long transient phase remains a critical problem to be solved.

The harmonic balance (HB) method directly overcomes this computational barrier by rigorously solving for the periodic state \cite{hall2002computation,hall2013harmonic,yan2023harmonic}. By expanding a periodic variable as a truncated Fourier series—or equivalently, sampling it at a finite set of time-domain collocation points—the physical time derivative is exactly mapped to a dense time-spectral operator that couples all temporal phases. This spectral transformation essentially converts the unsteady governing equations into an inherently quasi-steady system of coupled spatial equations, which can then be efficiently driven to convergence using pseudo-time relaxation or Newton-type algorithms. While HB and its time-spectral variants have achieved mature success in structural dynamics~\cite{labryer2009modeling}, turbomachinery aeroelasticity~\cite{wu2024efficient}, and periodic fluid flows~\cite{gopinath2005time}, their extension to RF plasma simulation is exceptionally promising. Not only is the discharge driven by a strictly prescribed external frequency, but both experimental diagnostics and baseline numerical responses of RF CCPs naturally exhibit strongly discrete harmonic components, making them an ideal physical system for frequency-domain or time-spectral resolution.

A recent study demonstrated the potential of HB for accelerating low-temperature plasma simulations by applying a high-dimensional HB formulation to a one-dimensional argon discharge \cite{zulevic2026accelerating}. While achieving substantial reductions in CPU time for density-only drift-diffusion-Poisson calculations, that benchmark work also brought to light the characteristic HB challenges of harmonic truncation and aliasing. However, to maintain numerical tractability, the plasma model in that study relied on the local-field approximation (LFA), wherein the electron energy equation is omitted and transport and ionization coefficients are assumed to be determined instantaneously by the local electric field. However, for realistic RF discharges, a physically rigorous description requires the local-mean-energy approximation (LMEA) to couple electron energy dynamics. Numerically, this coupling drastically escalates the stiffness and non-linearity of the system due to the introduction of energy transport, Joule heating, and strongly temperature-dependent reaction rates.

To address these limitations, this work develops a generalized time-domain HB framework for fluid simulations of RF CCPs based on the drift-diffusion approximation. The governing equations comprise species continuity and electron energy equations, tightly coupled with a phase-resolved Poisson equation. By employing a truncated Fourier expansion and evaluating the macroscopic variables at discrete temporal collocation points, the physical time derivatives are strictly transformed into time-spectral source terms. This spectral mapping inherently couples the transient states across all temporal collocation points, converting the original unsteady problem into a quasi-steady system and effectively bypassing the prohibitive computational overhead of conventional time-marching methods.

Solving this pseudo-spectral system, however, presents a significant computational challenge. The primary bottleneck stems from the numerical stiffness introduced by the dense phase-coupling nature of the time spectral operator. An explicit treatment of these source terms inevitably amplifies errors and triggers severe instabilities at large pseudo-time steps. To ensure robust numerical stability, we propose an implicit pseudo-time marching strategy based on a spatiotemporal operator-splitting strategy. In this approach, the spatial transport and volumetric reaction terms are advanced via a local implicit relaxation method independently at each temporal collocation point, while the stiff time spectral source term is treated implicitly through a cell-local dense matrix inversion. Concurrently, a semi-implicit correction is integrated into the phase-wise Poisson solver to overcome the strict dielectric relaxation limit. By restricting all implicit inversion operations strictly to the cell level, this decoupled approach entirely avoids the assembly of a massive global Jacobian matrix. This ensures high memory efficiency while significantly widening the stability margins against strong nonlinearity and stiffness. To rigorously assess the overall performance of the proposed framework, comprehensive benchmark evaluations are conducted against fully converged time-marching solutions.

The remainder of this paper is organized as follows. Section~\ref{sec:methodology} details the numerical methodology, including the physical governing equations, the time-domain HB formulation, and the proposed implicit pseudo-time relaxation method. Section~\ref{sec:results} presents a rigorous spatiotemporal validation of the HB solver, followed by a systematic assessment of its harmonic convergence and computational acceleration for representative RF CCP configurations. Finally, Section~\ref{sec:conclusion} draws the main conclusions of this work.

\section{Mathematical and Numerical Formulation}\label{sec:methodology}

\subsection{Governing Equations and Nondimensionalization}

The radio-frequency capacitively coupled plasma (RF CCP) is modeled based on a macroscopic continuum fluid framework. Given that the characteristic length scales of the reactor are significantly smaller than the electromagnetic wavelength of the RF drive, inductive effects can be negligible. Consequently, the self-consistent electric field $\mathbf{E} = -\nabla \phi$ is rigorously described by the electrostatic approximation, where the potential $\phi$ is governed by Poisson's equation:
\begin{equation}
-\nabla\cdot\left(\epsilon_0\nabla\phi\right)
=
e\left(n_i-n_e\right),
\label{eq:poisson_dim}
\end{equation}
where $n_e$ and $n_i$ denote the electron and positive ion number densities, $e$ is the elementary charge, and $\epsilon_0$ is the vacuum permittivity.

To resolve the fundamental discharge kinetics, the transport model tracks the spatiotemporal evolution of electrons, positive ions, and neutral metastable atoms (denoted by the subscript $*$). Under the highly collisional regime typical of such discharges, the macroscopic species fluxes are adequately captured by the drift-diffusion approximation. The corresponding continuity equations are formulated as:
\begin{equation}
\frac{\partial n_e}{\partial t}
+
\nabla\cdot\boldsymbol{\Gamma}_e
=
S_e,
\label{eq:electron_continuity}
\end{equation}
\begin{equation}
\frac{\partial n_i}{\partial t}
+
\nabla\cdot\boldsymbol{\Gamma}_i
=
S_i,
\label{eq:ion_continuity}
\end{equation}
\begin{equation}
\frac{\partial n_*}{\partial t}
+
\nabla\cdot\boldsymbol{\Gamma}_*
=
S_*.
\label{eq:meta_continuity}
\end{equation}

The transport fluxes $\boldsymbol{\Gamma}$ incorporate both electromigration driven by the macroscopic electric field and concentration-gradient-driven diffusion. Because the metastable species are electrically neutral, their transport reduces to pure diffusion. The respective fluxes are explicitly given by:
\begin{equation}
\boldsymbol{\Gamma}_e
=
-\mu_e n_e \mathbf{E}
-
D_e\nabla n_e, 
\label{eq:electron_flux}
\end{equation}
\begin{equation}
\boldsymbol{\Gamma}_i
=
\mu_i n_i \mathbf{E}
-
D_i\nabla n_i, 
\label{eq:ion_flux}
\end{equation}
\begin{equation}
\boldsymbol{\Gamma}_*
=
-D_*\nabla n_*,
\label{eq:meta_flux}
\end{equation}
where $\mu$ and $D$ represent the mobilities and diffusion coefficients of the respective species. 

The volumetric source terms ($S_e$, $S_i$, and $S_*$) represent the net production and depletion rates dictated by gas-phase chemical kinetics. For the charged species, the primary generation mechanisms include direct electron-impact ionization from the background ground state, stepwise ionization of metastables, and metastable pooling. Assuming singly charged ions, the electron and ion sources are strictly equivalent ($S_e = S_i$) and can be expanded as:
\begin{equation}
S_e
=
k_{\rm ion} N_g n_e 
+
k_{\rm si} n_* n_e
+
k_{\rm mp} n_*^2,
\label{eq:source_terms}
\end{equation}
where $N_g$ is the number  density of background neutral gas, and $k_{\rm ion}$, $k_{\rm si}$, and $k_{\rm mp}$ are the rate coefficients for direct ionization, stepwise ionization, and metastable pooling, respectively.

To accurately capture the non-local and non-equilibrium electron dynamics, the electron energy density $\varepsilon_e = \frac{3}{2} n_e k_B T_e$ (with $T_e$ denoting the electron temperature in $\mathrm{K}$) is governed by a conservative transport equation:
\begin{equation}
\frac{\partial \varepsilon_e}{\partial t}
+
\nabla\cdot\boldsymbol{\Gamma}_{\varepsilon}
=
\mathbf{J}_e\cdot\mathbf{E}
-
\sum_{j} \Delta\mathcal{E}_{j} S_{j}.
\label{eq:electron_energy}
\end{equation}

Within the drift-diffusion framework, the electron energy flux $\boldsymbol{\Gamma}_{\varepsilon}$ is modeled assuming standard electron-enthalpy closure:
\begin{equation}
\boldsymbol{\Gamma}_{\varepsilon}
=
-\frac{5}{3}\mu_e\varepsilon_e\mathbf{E}
-
\frac{5}{3}D_e\nabla\varepsilon_e.
\label{eq:energy_flux}
\end{equation}

On the right-hand side of Eq.~\eqref{eq:electron_energy}, the term $\mathbf{J}_e\cdot\mathbf{E}$ accounts for the local Joule heating, where $\mathbf{J}_e=-e\boldsymbol{\Gamma}_e$ is the electron current density. The summation term encompasses the aggregate inelastic collisional energy losses. For each $j$-th collision channel, $S_j$ represents its volumetric reaction rate, and $\Delta\mathcal{E}_j$ dictates the specific threshold energy loss. 

Because the rate coefficients for electron-impact reactions are acutely sensitive to the electron energy, they are typically parameterized as functions of the local mean electron energy. In the present macroscopic formulation, this mean energy is defined directly from the transported state variables as:
\begin{equation}
\bar{\varepsilon}_e
=
\frac{\varepsilon_e}{n_e}
=
\frac{3}{2} k_B T_e.
\end{equation}

To alleviate the severe numerical stiffness caused by the vast multi-scale disparities in physical quantities, the governing system is rigorously nondimensionalized. Defining $L_{\rm ref}$, $n_{\rm ref}$, $T_{e,\rm ref}$, and $\Phi_{\rm ref}$ as the chosen reference length, number density, electron temperature, and electrostatic potential, respectively, the reference electron thermal speed and characteristic timescale are derived as:
\begin{equation}
v_{\rm ref}
=
\left(
\frac{8 k_B T_{e,\rm ref}}{\pi m_e}
\right)^{1/2},
\label{eq:vref_tref}
\end{equation}
\begin{equation}
t_{\rm ref}
=
\frac{L_{\rm ref}}{v_{\rm ref}},
\label{eq:tref}
\end{equation}
where $k_B$ is the Boltzmann constant and $m_e$ is the electron mass. The reference electron energy density is correspondingly scaled by:
\begin{equation}
\varepsilon_{\rm ref}
=
\frac{3}{2}
n_{\rm ref} k_B T_{e,\rm ref}.
\label{eq:eref}
\end{equation}

All subsequent secondary variables—including transport coefficients, reaction rate coefficients, and vacuum permittivity—are scaled consistently against these primary reference values. This systematic scaling ensures that the dimensionless governing equations maintain the exact mathematical structure of their dimensional counterparts; thus, the dimensionless forms are omitted here for brevity.

\subsection{Time-Domain Harmonic Balance Formulation}
\label{subsec:hb_formulation}

The RF-periodic solution is assumed to reach a quasi-steady state with the fundamental period
\begin{equation}
    T=\frac{2\pi}{\omega}.
\end{equation}

For a scalar periodic variable $Q(\mathbf{x},t)$, the representation based on a truncated Fourier series with $N_H$ harmonics is
\begin{equation}
    Q(\mathbf{x},t)
    \approx
    \sum_{k=-N_H}^{N_H}
    \widehat{Q}_k(\mathbf{x})e^{ik\omega t},
    \label{eq:complex_fourier_series}
\end{equation}
where $\widehat{Q}_k$ represents the complex Fourier coefficients satisfying the conjugate symmetry $\widehat{Q}_{-k}=\widehat{Q}_{k}^{*}$. The number of temporal collocation points is chosen to satisfy the Nyquist criterion:
\begin{equation}
    N_T=2N_H+1,
\end{equation}
with the temporal collocation points uniformly distributed over one fundamental period:
\begin{equation}
    t_m=\frac{mT}{N_T},
    \qquad
    m=0,1,\ldots,N_T-1 .
    \label{eq:phase_points}
\end{equation}

By assembling the flow variables at these collocation points into time-domain vectors, e.g., $\mathbf{Q} = [\mathbf{Q}|_{t_0}, \dots, \mathbf{Q}|_{t_{2N_\text{H}}}]^T$, the discrete Fourier transform (DFT) provides a direct mapping between the solution values at the discrete sub-time levels and their corresponding frequency-domain Fourier coefficients:
\begin{equation}
\widehat{\mathbf{Q}} = \mathbf{D}\,  \mathbf{Q}.
\label{eq:dft_Q}
\end{equation}

The discrete Fourier transform (DFT) matrix $\mathbf{D}$ and its inverse $\mathbf{D}^{-1}$ govern this transformation. Specifically, the inverse DFT matrix $\mathbf{D}^{-1}$ is explicitly defined using the complex exponential basis functions evaluated at the collocation points as follows:
\begin{equation}
\mathbf{D}^{-1} =
\begin{bmatrix}
1 & e^{i\omega t_0} & e^{i 2\omega t_0} & \dots & e^{-i\omega t_0} \\
1 & e^{i\omega t_1} & e^{i 2\omega t_1} & \dots & e^{-i\omega t_1} \\
\vdots & \vdots & \vdots & \ddots & \vdots \\
1 & e^{i\omega t_{N_T-1}} & e^{i 2\omega t_{N_T-1}} & \dots & e^{-i\omega t_{N_T-1}}
\end{bmatrix}.
\label{eq:D_matrix}
\end{equation}

Consequently, the differentiation in physical time can be exactly represented at the discrete temporal collocation points by a matrix multiplication:
\begin{equation}
    \frac{\partial \mathbf{Q}}{\partial t}
    =
    \mathbf{D}^{-1}\mathbf{K}\widehat{\mathbf{Q}}
    = \mathbf{D}^{-1}\mathbf{K}\mathbf{D}\,  \mathbf{Q} =
    \mathbf{E}\mathbf{Q},
    \label{eq:hb_derivative_matrix}
\end{equation}
where matrix $\mathbf{K}$ is defined as
\begin{equation}
    \mathbf{K}
    =
    i\omega
    \operatorname{diag}(0,1,\ldots,N_H,-N_H,\ldots,-1).
\end{equation}

Although Eq.~\eqref{eq:hb_derivative_matrix} is mathematically derived using a complex Fourier basis, the resulting operator $\mathbf{E}$ analytically maps real discrete-time vectors to real temporal derivatives. For an odd number of collocation points $N_T$, this operator can be written explicitly as a real, dense, skew-symmetric matrix:
\begin{equation}
    (\mathbf{E})_{m\ell}
    =
    \begin{cases}
    \displaystyle
    \frac{\omega}{2}
    (-1)^{m-\ell}
    \csc\left[\frac{\pi(m-\ell)}{N_T}\right],
    & m\neq \ell, \\
    0, & m=\ell .
    \end{cases}
    \label{eq:explicit_time_spectral_operator}
\end{equation}

The strict skew-symmetry of $\mathbf{E}$ intrinsically reflects the non-dissipative character of the time spectral operator within the retained harmonic subspace. Any harmonic truncation errors and nonlinear aliasing effects are systematically controlled by increasing $N_H$, which will be assessed in the numerical results.

By applying the time spectral operator defined in Eq.~\eqref{eq:hb_derivative_matrix} to the continuous governing equations, the continuous time derivatives are replaced by discrete couplings across all phases. Let $\mathbf{n}_e^{*}$, $\mathbf{n}_i^{*}$, $\mathbf{n}_*^{*}$, and $\boldsymbol{\varepsilon}_e^{*}$ denote the HB phase vectors of the transported variables at a fixed spatial location. For a specific temporal collocation point $t_m$, the HB form of the plasma fluid system is written as:
\begin{subequations}
\label{eq:hb_continuous_phase}
\begin{align}
    \sum_{\ell=0}^{N_T-1}
    (\mathbf{E})_{m\ell}n_{e,\ell}
    +\nabla\cdot\boldsymbol{\Gamma}_{e,m}
    &=
    S_{e,m}, \\
    \sum_{\ell=0}^{N_T-1}
    (\mathbf{E})_{m\ell}n_{i,\ell}
    +\nabla\cdot\boldsymbol{\Gamma}_{i,m}
    &=
    S_{i,m}, \\
    \sum_{\ell=0}^{N_T-1}
    (\mathbf{E})_{m\ell}n_{*,\ell}
    +\nabla\cdot\boldsymbol{\Gamma}_{*,m}
    &=
    S_{*,m}, \\
    \sum_{\ell=0}^{N_T-1}
    (\mathbf{E})_{m\ell}\varepsilon_{e,\ell}
    +\nabla\cdot\boldsymbol{\Gamma}_{\varepsilon,m}
    &=
    P_{J,m}
    -
    \sum_{j} \Delta\mathcal{E}_{j} S_{j,m}.
\end{align}
\end{subequations}

Crucially, all strongly nonlinear transport coefficients, volumetric reaction rates (e.g., $S_{e,m}$, $S_{*,m}$), Joule heating terms ($P_{J,m}$), and boundary fluxes are evaluated directly at the temporal collocation point $t_m$. This decoupling is the central advantage of the time-domain HB implementation. By evaluating the highly non-linear plasma kinetics and physical constraints strictly locally in each temporal collocation point, the method completely avoids frequency-domain convolutions, while the periodic time evolution remains globally and rigorously governed by the dense time spectral operator $\mathbf{E}$.

Furthermore, because Poisson's equation contains no explicit physical time derivative, it effectively acts as an elliptic constraint evaluated independently at each temporal collocation point:
\begin{equation}
    \nabla\cdot(\epsilon_0\nabla\phi_m)
    =
    -e(n_{i,m}-n_{e,m}),
    \qquad
    m=0,1,\ldots,N_T-1 .
    \label{eq:hb_poisson_phase}
\end{equation}

The corresponding time-varying boundary conditions are evaluated at these temporal collocation points, ensuring that the electrostatic constraints are perfectly synchronized with the HB temporal collocation.

\subsection{Finite-Volume Spatial Discretization}
\label{subsec:finite_volume_hb}

To numerically resolve the RF-periodic discharge dynamics, the governing equations in the HB form must be discretized spatially. The computational domain is discretized by non-overlapping cell-centered control volumes. Integrating the governing system over a control volume $\Omega_i$ with volume $V_i$ yields the discrete nonlinear HB residual for cell $i$ at a specific temporal collocation point $t_m$ (denoted by the index $m$):
\begin{equation}
    \mathbf{G}_{i,m}
    =
    - \sum_{f\in\partial\Omega_i} A_f\mathbf{F}_{f,m}
    + V_i\mathbf{S}_{i,m}
    - V_i \sum_{\ell=0}^{N_T-1} \mathbf{E}_{m\ell}\mathbf{Q}_{i,\ell}
    = \mathbf{0} ,
    \label{eq:fv_hb_residual_local}
\end{equation}
where $A_f$ represents the face area, and $\mathbf{E}_{m\ell}$ denotes the components of the dense $N_T \times N_T$ time spectral operator $\mathbf{E}$. 

At a given collocation point $m$, the conservative variable vector $\mathbf{Q}_{i,m}$ and the corresponding volumetric source vector $\mathbf{S}_{i,m}$ for cell $i$ are compactly grouped as:
\begin{equation}
    \mathbf{Q}_{i,m} = \begin{bmatrix} n_{e,i,m} & n_{i,i,m} & n_{*,i,m} & \varepsilon_{e,i,m} \end{bmatrix}^T ,
\end{equation}
\begin{equation}
    \mathbf{S}_{i,m} = \begin{bmatrix} S_{e,i,m} & S_{i,i,m} & S_{*,i,m} & P_{J,i,m} - \sum_{j} \Delta\mathcal{E}_{j} S_{j,i,m} \end{bmatrix}^T .
\end{equation}
in which the volumetric electrostatic work is evaluated in a face-consistent discrete form:
\begin{equation}
    P_{J,i,m}V_i
    \approx
    - \phi_{i,m} \sum_{f\in\partial\Omega_i} A_f J_{e,f,m} ,
    \label{eq:discrete_joule}
\end{equation}
where $J_{e,f,m} = -e(\boldsymbol{\Gamma}_{e,f,m}\cdot\mathbf{n}_f)$ is the local electron-current contribution passing through face $f$.

As shown in Eq.~\ref{eq:fv_hb_residual_local}, the total residual $\mathbf{G}_{i,m}$ is inherently composed of a purely physical residual and a time spectral source term. The physical spatial-chemical residual, denoted as $\mathbf{R}_{i,m}^{\text{phys}}$, is defined as:
\begin{equation}
    \mathbf{R}_{i,m}^{\text{phys}} 
    = 
    - \sum_{f\in\partial\Omega_i} A_f\mathbf{F}_{f,m} 
    + V_i\mathbf{S}_{i,m} ,
    \label{eq:physical_residual}
\end{equation}
which encapsulates the flux and reactive source evaluations at the temporal collocation point $t_m$. Conversely, the remaining term $V_i \sum_{\ell} \mathbf{E}_{m\ell}\mathbf{Q}_{i,\ell}$ represents the time spectral source term introduced by the time-domain HB. It mathematically couples the local state at the current temporal collocation point to the states at all other temporal collocation points, thereby  driving the transient evolution toward a dynamic steady state.

A fundamental advantage of this time-domain HB framework is that the physical residual $\mathbf{R}_{i,m}^{\text{phys}}$ is evaluated entirely independently at each temporal collocation point. Because the spatial non-linear fluxes $\mathbf{F}_{f,m}$ and chemical sources are temporally dependent, the formulation strictly follows the identical reconstruction and evaluation procedures used in time-marching solvers. By using only the  macroscopic variables at $t_m$, the method completely avoids the complex frequency convolutions typically encountered in frequency-domain methods.

At convergence, the algebraic residual system $\mathbf{G}_{i,m} = \mathbf{0}$ is strictly satisfied for every spatial cell across all temporal collocation points, yielding the spectrally accurate RF-periodic steady state.

\subsection{Implicit Spatio-Temporal Decoupling and Block Relaxation}
\label{subsec:implicit_solver}

To efficiently drive the discrete residual system to convergence, an implicit pseudo-time integration method is introduced by defining a pseudo-time parameter $\tau$. Let $s$ denote the pseudo-time iteration index, and then define the increment as 
\begin{equation}
    \Delta\mathbf{Q} = \mathbf{Q}^{s+1} - \mathbf{Q}^s.
\end{equation}

If a standard  implicit backward Euler linearization were applied directly to Eq.~\eqref{eq:fv_hb_residual_local}, the corrections of all $N_T$ collocation points within a cell would be monolithically coupled by the time spectral operator $\mathbf{E}$. To mathematically represent this monolithic system at cell $i$, we temporarily concatenate the conservative variables, physical spatial-chemical residual and total residuals across all temporal collocation points into augmented vectors:
\begin{equation}
    \mathbf{Q}_i = \begin{bmatrix} \mathbf{Q}_{i,0} & \dots & \mathbf{Q}_{i,N_T-1} \end{bmatrix}^T,
\end{equation}
\begin{equation}
    \mathbf{R}_i^{\text{phys}} = \begin{bmatrix} \mathbf{R}_{i,0}^{\text{phys}} & \dots & \mathbf{R}_{i,N_T-1} ^{\text{phys}}\end{bmatrix}^T .
\end{equation}
\begin{equation}
    \mathbf{G}_i = \begin{bmatrix} \mathbf{G}_{i,0} & \dots & \mathbf{G}_{i,N_T-1} \end{bmatrix}^T .
\end{equation}

The unfactored implicit linear system for cell $i$ then reads:
\begin{equation}
    \left[ 
    \frac{V_i}{\Delta\tau} \mathbf{I} 
    + V_i (\mathbf{E} \otimes \mathbf{I}_q) 
    - \frac{\partial\mathbf{R}_i^{\text{phys}}}{\partial\mathbf{Q}_i} 
    \right] \Delta\mathbf{Q}_i 
    - \sum_{j\in\mathcal{N}(i)} 
    \frac{\partial\mathbf{R}_i^{\text{phys}}}{\partial\mathbf{Q}_j} \Delta\mathbf{Q}_j 
    = \mathbf{G}_i(\mathbf{Q}^s, \phi^s) ,
    \label{eq:linear_system_monolithic}
\end{equation}
where $\mathcal{N}(i)$ is the set of neighboring spatial cells, $\mathbf{I}_q$ is the $N_q \times N_q$ identity matrix corresponding to the $N_q$ conservative variables and $\otimes$ denotes the Kronecker product. 

To avoid the steep computational expense associated with inverting the fully coupled local augmented block, a spatio-temporal approximate factorization (AF) is introduced. By defining the local spatial-chemical Jacobian at a specific temporal collocation point $m$ as 

\begin{equation}
\mathbf{A}_{ii,m} = -\frac{1}{V_i} \frac{\partial\mathbf{R}_{i,m}^{\text{phys}}}{\partial\mathbf{Q}_{i,m}},
\end{equation}
which inherently encapsulates both the exact spatial flux derivatives and the stiff chemical source Jacobians and utilizing the multi-time block form
\begin{equation}
\mathbf{A}_{ii} = \operatorname{diag}(\mathbf{A}_{ii,0}, \dots, \mathbf{A}_{ii,N_T-1}),
\end{equation}
the local implicit coefficient matrix can be algebraically expanded as:
\begin{equation}
    \mathbf{I} + \Delta\tau (\mathbf{E} \otimes \mathbf{I}_q) + \Delta\tau \mathbf{A}_{ii} 
    = 
    \big(\mathbf{I} + \Delta\tau (\mathbf{E} \otimes \mathbf{I}_q)\big) \big(\mathbf{I} + \Delta\tau \mathbf{A}_{ii}\big) 
    - (\Delta\tau)^2 (\mathbf{E} \otimes \mathbf{I}_q) \mathbf{A}_{ii} .
    \label{eq:exact_expansion}
\end{equation}

Because the last term on the right-hand side of the above equation is of second order in pseudo-time, it introduces minimal factorization error during the iterative relaxation. More importantly, since it operates directly on the state increment $\Delta\mathbf{Q}_i$, this term strictly vanishes as the system converges toward the exact RF-periodic steady state. Therefore, it can be safely neglected without compromising the final converged solution, successfully decoupling the local implicit coefficient matrix into a product of distinct temporal and spatial structures:
\begin{equation}
    \mathbf{I} + \Delta\tau (\mathbf{E} \otimes \mathbf{I}_q) + \Delta\tau \mathbf{A}_{ii} 
    \approx 
    \big(\mathbf{I} + \Delta\tau (\mathbf{E} \otimes \mathbf{I}_q)\big) \big(\mathbf{I} + \Delta\tau \mathbf{A}_{ii}\big) .
    \label{eq:approximate_factorization}
\end{equation}

\begin{figure}[h]
    \centering
    \includegraphics[width=0.8\textwidth]{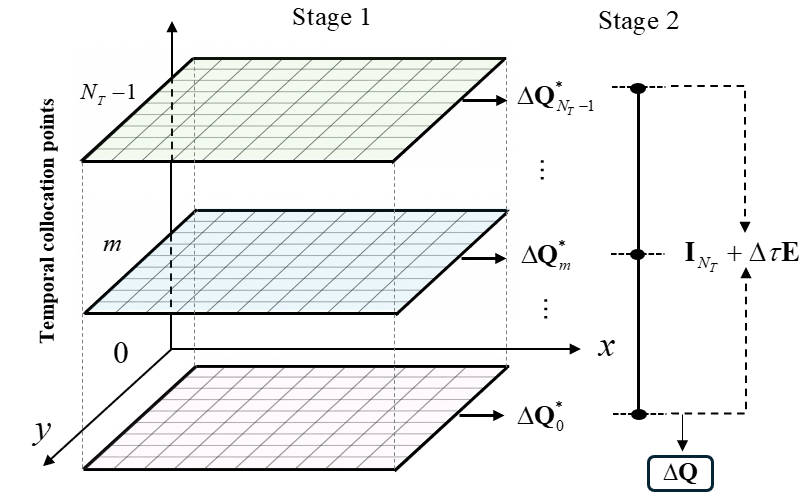}
    \caption{Schematic illustration of the spatio-temporal decoupling strategy in the time-domain harmonic balance solver.}
    \label{fig:hb_spacetime_decoupling}
\end{figure}

Through this approximate factorization, the monolithic system can be decomposed into two sequential and decoupled relaxation stages executed at each pseudo-time step, as schematically illustrated in Fig.~\ref{fig:hb_spacetime_decoupling}.

\paragraph{Stage 1: Spatial Block-Implicit Relaxation}
The first stage evaluates the spatial operator to solve the spatial and chemical coupling, yielding an intermediate state correction $\Delta\mathbf{Q}^*$:
\begin{equation}
    \big(\mathbf{I} + \Delta\tau \mathbf{A}_{ii}\big) \Delta\mathbf{Q}_i^* - \frac{\Delta\tau}{V_i} \sum_{j\in\mathcal{N}(i)} \frac{\partial\mathbf{R}_i^{\text{phys}}}{\partial\mathbf{Q}_j} \Delta\mathbf{Q}_j^* = \frac{\Delta\tau}{V_i} \mathbf{G}_i^s .
\end{equation}

Crucially, because the spatial-chemical Jacobian $\mathbf{A}_{ii}$ involves no temporal cross-coupling, all temporal collocation points are completely decoupled in this stage. This allows the local spatial system to be solved independently for each temporal collocation point $m$ via a relaxation sweep. For cell $i$ at a specific collocation point $m$, the local update equations during the forward and backward spatial sweeps are reduced to:
\begin{equation}
    D_{i,m} \Delta\mathbf{Q}_{i,m}^{**} = \frac{\Delta\tau}{V_i} \mathbf{G}_{i,m}^s - \frac{\Delta\tau}{V_i} \sum_{j<i} L_{ij,m} \Delta\mathbf{Q}_{j,m}^{**} ,
\end{equation}
\begin{equation}
    D_{i,m} \Delta\mathbf{Q}_{i,m}^* = D_{i,m} \Delta\mathbf{Q}_{i,m}^{**} - \frac{\Delta\tau}{V_i} \sum_{j>i} U_{ij,m} \Delta\mathbf{Q}_{j,m}^* ,
\end{equation}
where $L$ and $U$ are the exact spatial lower and upper neighbor Jacobian matrices. 

To avoid the high computational and storage costs of assembling exact spatial flux Jacobians while maintaining strong diagonal dominance, the exact local operator $\mathbf{A}_{ii,m}$ is robustly approximated. The resulting local diagonal preconditioning block $D_{i,m}$ is constructed as a compact $N_q \times N_q$ matrix:
\begin{equation}
    D_{i,m} = \mathbf{I}_q + \Delta\tau \operatorname{diag}(\lambda_{e,i,m}, \lambda_{i,i,m}, \lambda_{*,i,m}, \lambda_{\varepsilon,i,m}) - \Delta\tau \mathbf{J}_{{\rm chem},i,m} + \mathbf{D}_{{\rm damp},i,m} .
\end{equation}

The transport spectral radii $\lambda_{s,i,m}$ are conservatively evaluated over all boundary faces $f \in \partial\Omega_i$ based on local drift-diffusion velocities:
\begin{subequations}
\label{eq:spectral_radii}
\begin{align}
    \lambda_{s,i,m} &\approx \sum_{f\in\partial\Omega_i}
    \frac{A_f}{V_i}
    \left( \frac{1}{2}\mu_s |\mathbf{E}_{f,m}| + c_D\frac{D_q}{|\Delta \mathbf{r}_f|} \right), \quad (q = e, i, *) ,\\
    \lambda_{\varepsilon,i,m} &\approx \sum_{f\in\partial\Omega_i}
    \frac{A_f}{V_i}
    \left( \frac{5}{6}\mu_e |\mathbf{E}_{f,m}| + \frac{5}{3}c_D\frac{D_e}{|\Delta \mathbf{r}_f|} \right) ,
\end{align}
\end{subequations}
where $c_D$ is a numerical diffusion scaling constant, and $|\Delta \mathbf{r}_f|$ is the characteristic distance between neighboring cell centers. Concurrently, to eliminate severe numerical noise and oscillatory derivatives during the relaxation sweeps, the highly sensitive reaction rate coefficients (e.g., $k_{\rm ion}, k_{\rm exc}$) are kept frozen as local constants during the analytical differentiation of the source vector. For the specific multi-moment plasma system, the symbolic structure of this frozen-rate chemistry Jacobian block is formulated as:

\begin{equation}
    \mathbf{J}_{{\rm chem},i,m}
    =
    \left.
    \frac{\partial \mathbf{S}_{i,m}}{\partial \mathbf{Q}_{i,m}}
    \right|_{\text{frozen } k}
    =
    \begin{bmatrix}
        \frac{\partial S_{e}}{\partial n_{e}} & \frac{\partial S_{e}}{\partial n_{i}} & \frac{\partial S_{e}}{\partial n_{*}} & \frac{\partial S_{e}}{\partial \varepsilon_{e}} \\
        \frac{\partial S_{i}}{\partial n_{e}} & \frac{\partial S_{i}}{\partial n_{i}} & \frac{\partial S_{i}}{\partial n_{*}} & \frac{\partial S_{i}}{\partial \varepsilon_{e}} \\
        \frac{\partial S_{*}}{\partial n_{e}} & \frac{\partial S_{*}}{\partial n_{i}} & \frac{\partial S_{*}}{\partial n_{*}} & \frac{\partial S_{*}}{\partial \varepsilon_{e}} \\
        \frac{\partial S_{\varepsilon}}{\partial n_{e}} & \frac{\partial S_{\varepsilon}}{\partial n_{i}} & \frac{\partial S_{\varepsilon}}{\partial n_{*}} & \frac{\partial S_{\varepsilon}}{\partial \varepsilon_{e}}
    \end{bmatrix}_{i,m, \text{frozen } k} ,
    \label{eq:chemistry_jacobian_matrix}
\end{equation}
where the partial derivatives of the source vector with respect to conservative variables are analytically evaluated. Under this frozen-rate assumption, representative non-zero Jacobian entries for the electron volumetric source and energy sink terms elegantly simplify to:
\begin{subequations}
\label{eq:approx_source_derivatives}
\begin{align}
    \frac{\partial S_{e,i,m}}{\partial n_{e,i,m}}
    &\approx
    k_{\rm ion} N_g + k_{\rm si} n_{*,i,m} , \\
    \frac{\partial S_{\varepsilon,i,m}}{\partial n_{e,i,m}}
    &\approx
    - \sum_{j} \Delta\mathcal{E}_{j} \left( \left. \frac{\partial S_{j,i,m}}{\partial n_{e,i,m}} \right|_{\text{frozen } k} \right) .
\end{align}
\end{subequations}

The remaining non-zero elements in $\mathbf{J}_{{\rm chem},i,m}$ are derived in an identical analytical manner. To enforce diagonal dominance and guarantee numerical stability under extreme chemical transients, a diagonal damping matrix $\mathbf{D}_{{\rm damp},i,m}$ is explicitly added, whose entries are proportional to the absolute row sums of $\mathbf{J}_{{\rm chem},i,m}$.

\paragraph{Stage 2: Variable-Decoupled Temporal Relaxation}
Once the spatial intermediate correction $\Delta\mathbf{Q}_i^*$ is obtained, the second stage accounts for the dense temporal coupling by solving the remaining temporal factor within each independent control volume:
\begin{equation}
    \big(\mathbf{I} + \Delta\tau (\mathbf{E} \otimes \mathbf{I}_q)\big) \Delta\mathbf{Q}_i = \Delta\mathbf{Q}_i^* .
\end{equation}

A pivotal computational attribute of this step is that the time spectral operator $\mathbf{E}$ acts exclusively across the temporal collocation points and introduces no cross-coupling between different fluid variables. Consequently, this multi-variable temporal block system can be segregated into $N_q$ independent scalar systems. For each distinct fluid variable $q \in \{n_e, n_i, n_*, \varepsilon_e\}$, the time spectral coupling is resolved by a localized dense matrix inversion:
\begin{equation}
    \left(\mathbf{I}_{N_T} + \Delta\tau \mathbf{E}\right) \Delta\mathbf{Q}_{i, q} = \Delta\mathbf{Q}_{i, q}^* ,
\end{equation}
where $\mathbf{I}_{N_T}$ is the $N_T \times N_T$ identity matrix. This variable-by-variable segregation drops the local temporal inversion cost from a monolithic $\mathcal{O}((N_q N_T)^3)$ down to $N_q \times \mathcal{O}(N_T^3)$. Finally, the local physical state is advanced using a dynamic relaxation limiter to guarantee strictly positive species densities and energies:
\begin{equation}
    \mathbf{Q}_i^{s+1} = \mathbf{Q}_i^s + \alpha \Delta\mathbf{Q}_i ,
    \label{eq:state_update}
\end{equation}
where $\alpha \in (0, 1]$ is a specified damping factor to prevent non-physical solutions.

\subsection{Temporally Decoupled Semi-Implicit Poisson Update}
\label{subsec:semiimplicit_poisson}

Solving Poisson's equation decoupled from the charged-particle transport often imposes a severe dielectric-relaxation pseudo-time step limit, leading to weak electrostatic coupling and numerical instability. To alleviate this issue and ensure robust convergence within the time-domain HB framework, we employ a semi-implicit correction, following the approaches established in \cite{kushner2009hybrid,li2025fast,ventzek1993two}. This method anticipates the fast electron number density response to potential variations over the pseudo-time step $\Delta\tau_i$.

To construct this semi-implicit operator, the implicit treatment is exclusively applied to the electron drift term during the prediction step. Concurrently, the electron diffusion is evaluated explicitly, and the chemical reaction source terms are neglected. Let $s$ denote the pseudo-time iteration step. By evaluating the mobility and electron density at the known iteration step $s$, the implicit drift flux at a specific temporal collocation point $m$ is approximated as 
\begin{equation}
 \boldsymbol{\Gamma}_{e,m}^{s+1} \approx -\mu_e n_{e,m}^s \mathbf{E}_m^{s+1} = \mu_e n_{e,m}^s \nabla\phi_m^{s+1}.   
\end{equation}

Substituting this predicted flux  into the electron continuity equation and subsequently into Poisson's equation yields the augmented semi-implicit continuous operator:
\begin{equation}
    -\nabla\cdot\left(\epsilon_0\nabla\phi_m^{s+1}\right)
    - \nabla\cdot\left(e\mu_e n_{e,m}^s\Delta\tau_i\nabla\phi_m^{s+1}\right)
    = e(n_{i,m}^s-n_{e,m}^s)+\mathcal{D}_{e,m}^s,
    \label{eq:semi_poisson_continuous}
\end{equation}
where $\mathcal{D}_{e,m}^s = -e \Delta\tau_i \nabla\cdot\left(D_e \nabla n_{e,m}^s\right)$ represents the explicit electron diffusion contribution.

A unique property of the electrostatic Poisson's equation in the HB formulation is the inherent absence of a physical time derivative. Consequently, the time spectral operator $\mathbf{E}$ does not couple the electric potential across different temporal collocation points. The Poisson system is naturally block-diagonal, allowing the elliptic equation to be solved completely independently at each temporal collocation point $m$. 

Integrating Eq.~\eqref{eq:semi_poisson_continuous} over a control volume $\Omega_i$ yields the discrete, decoupled finite-volume update for cell $i$ at a specific temporal collocation point $m$:
\begin{align}
    \sum_{f\in\partial\Omega_i}
    &
    \epsilon_{{\rm eff},f,m}
    \frac{A_f}{|\Delta \mathbf{r}_f|}
    \left(
    \phi_{i,m}^{s+1}
    -
    \phi_{nb,m}^{s+1}
    \right)
    \notag\\
    &=
    e(n_{i,i,m}^s-n_{e,i,m}^s)V_i
    -
    e\Delta\tau_i D_e
    \sum_{f\in\partial\Omega_i}
    A_f(\nabla n_e)_{f,m}^s\cdot\mathbf{n}_f ,
    \label{eq:semiimplicit_poisson_discrete}
\end{align}
where subscript $nb$ denotes the adjacent neighbor cell (or boundary face) sharing face $f$, and $|\Delta \mathbf{r}_f|$ is the characteristic distance. To maintain accuracy on distorted elements, the face-normal potential gradients in Eq.~\eqref{eq:semiimplicit_poisson_discrete} are further corrected by incorporating non-orthogonal cross-diffusion terms. 

Through this formulation, the original Laplacian operator is mathematically augmented by an effective face permittivity:
\begin{equation}
    \epsilon_{{\rm eff},f,m}
    =
    \epsilon_0
    +
    e\mu_e n_{e,f,m}^s\Delta\tau_i .
\end{equation}

In the algorithmic workflow, the electric potential is not advanced simultaneously with the fluid variables. Instead, this decoupled Poisson system is solved immediately after the macroscopic variables are updated via the spatial and temporal relaxation processes (Stages 1 and 2). Once the new potential $\phi_m^{s+1}$ is obtained independently across all temporal collocation points, the local electric field is strictly recomputed before evaluating the transport fluxes for the next pseudo-time step.

\subsection{Boundary Conditions}
\label{sec:boundary_conditions}

To complete the physical model and close the semi-discrete finite-volume system, physically consistent boundary conditions must be specified at the solid electrodes. In the present formulation, the electrostatic potential is imposed via Dirichlet boundary conditions, while the transport equations for charged particles and electron energy are closed through wall-flux boundary conditions. All boundary states are evaluated independently at each temporal collocation point $m$.

For the present RF capacitively coupled discharge, the powered electrode is driven by a prescribed sinusoidal voltage. At a specific temporal collocation point $t_m$, the boundary potential is given by:
\begin{equation}
    \phi_{{\rm p},m} = V_{\rm RF}\sin\left(2\pi f_{\rm RF} t_m +\varphi_0\right),
    \label{eq:rf_potential_bc}
\end{equation}
where $V_{\rm RF}$ is the peak voltage amplitude, $f_{\rm RF}$ is the driving frequency, and $\varphi_0$ is the phase shift. The grounded electrode is fixed at a constant reference potential 
\begin{equation}
   \phi_{{\rm g},m} = 0. 
\end{equation}

Driven predominantly by the strong sheath electric field, ions are accelerated toward the wall. The theoretical drift-driven ion flux at the boundary is given by:
\begin{equation}
    \boldsymbol{\Gamma}_{i,m}\cdot\mathbf{n}_w = \mu_i n_{i,m} \mathbf{E}_m\cdot\mathbf{n}_w,
    \label{eq:ion_wall_bc_theory}
\end{equation}
where $\mathbf{n}_w$ is the outward unit normal vector pointing from the plasma domain to the wall. In the actual numerical implementation, to prevent unphysical inward ion drift during highly transient iterative steps, a pure upwind condition is enforced by limiting this flux to non-negative values, i.e., 
\begin{equation}
    \boldsymbol{\Gamma}_{i,m}\cdot\mathbf{n}_w = \max(\boldsymbol{\Gamma}_{i,m}\cdot\mathbf{n}_w, 0). 
\end{equation}

For the electron and electron energy fluxes, the physical treatment at the wall depends heavily on the thermal assumptions. In this work, two distinct sets of flux boundary conditions are implemented within the time-spectral framework to accommodate different benchmarking requirements:

\paragraph{1. Prescribed Wall Temperature}
The first set assumes an energy-independent electron flux, typically associated with a prescribed constant electron temperature at the wall \cite{lymberopoulos1993fluid}. The macroscopic electron loss is modeled using a constant surface recombination velocity:
\begin{equation}
    \boldsymbol{\Gamma}_{e,m}\cdot\mathbf{n}_w = k_s n_{e,m} - \gamma (\boldsymbol{\Gamma}_{i,m}\cdot\mathbf{n}_w),
    \label{eq:electron_wall_bc_indep}
\end{equation}
where $k_s$ is the effective electron surface recombination coefficient (assuming a unity sticking coefficient) and $\gamma$ is the secondary electron emission (SEE) coefficient. Because the flux does not depend on the local energy, the electron energy equation must be closed by explicitly specifying a Dirichlet boundary condition based on the prescribed wall temperature $T_{e,w}$:
\begin{equation}
    \varepsilon_{e,w,m} = \frac{3}{2} n_{e,m} k_B T_{e,w}.
    \label{eq:wall_energy_bc}
\end{equation}

\paragraph{2. Zero-Gradient Temperature Extrapolation}
The second set introduces electron energy dependence derived from kinetic theory, assuming the electron temperature at the boundary is extrapolated from the adjacent interior cell (i.e., a zero-gradient condition for $T_e$) \cite{sakiyama2006corona}. The macroscopic electron loss rate is governed by the thermal flux of a half-Maxwellian distribution:
\begin{equation}
    \boldsymbol{\Gamma}_{e,m}\cdot\mathbf{n}_w = \frac{1}{4} n_{e,m} \left( \frac{8 k_B T_{e,m}}{\pi m_e} \right)^{1/2} - \gamma (\boldsymbol{\Gamma}_{i,m}\cdot\mathbf{n}_w),
    \label{eq:electron_wall_bc_dep}
\end{equation}
where $T_{e,m}$ is the locally extrapolated electron temperature, and $m_e$ is the electron mass. Consistent with this particle flux, the transported electrons carry their respective mean thermal energies to the wall. The corresponding electron energy boundary flux is robustly evaluated as:
\begin{equation}
    \boldsymbol{\Gamma}_{\varepsilon,m}\cdot\mathbf{n}_w = \frac{5}{3} \bar{\varepsilon}_{e,m} \left[ \frac{1}{4} n_{e,m} \left( \frac{8 k_B T_{e,m}}{\pi m_e} \right)^{1/2} - \gamma (\boldsymbol{\Gamma}_{i,m}\cdot\mathbf{n}_w) \right].
    \label{eq:energy_wall_flux_bc}
\end{equation}

\section{Numerical Setups and Results}\label{sec:results}

To comprehensively evaluate the solution accuracy and computational efficiency of the proposed time-domain HB method, a representative RF-CCP benchmark is investigated. The computed macroscopic fluid solutions are systematically compared against both a well-established reference solution and the time-marching results performed within the same unified finite-volume framework.

\subsection{Physical Model and Numerical Configuration}

\begin{figure}[htbp]
    \centering
    \includegraphics[width=0.8\textwidth]{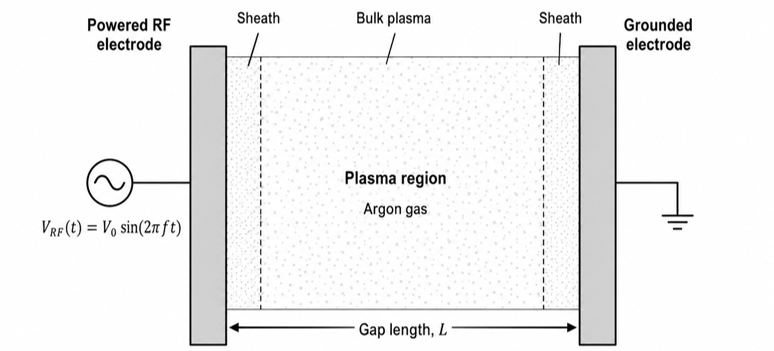}
    \caption{Schematic diagram of the one-dimensional RF CCP discharge configuration.}
    \label{fig:schematic_diagram}
\end{figure}

The proposed time-domain HB method is evaluated using a standard one-dimensional, parallel-plate RF CCP benchmark. A schematic diagram of the discharge configuration is illustrated in Fig.~\ref{fig:schematic_diagram}. The plasma reactor is bounded by a powered RF electrode on the left and a grounded electrode on the right, filled with neutral argon gas. As depicted, the discharge gap is macroscopically characterized by a central quasi-neutral bulk plasma region enclosed by two oscillating space-charge sheaths adjacent to the electrodes. Because the transverse dimensions of the electrodes are assumed to be infinitely large compared to the gap distance, radial edge effects can be safely neglected. Consequently, the plasma distribution can be treated as one-dimensional along the axial direction. 

The geometric parameters and macroscopic operating conditions are summarized in Table~\ref{tab:operating_conditions}. 
\begin{table}[h]
    \centering
    \caption{Geometric parameters and operating conditions for the argon CCP benchmark.}
    \label{tab:operating_conditions}
    \begin{tabular}{llll}
        \toprule
        \textbf{Parameter} & \textbf{Symbol} & \textbf{Unit} & \textbf{Value} \\
        \midrule
        Interelectrode distance      & $L$       & $\mathrm{m}$               & $0.0254$ \\
        Background gas pressure      & $p$         & $\mathrm{Torr}$            & $1$ \\
        Gas temperature              & $T_g$       & $\mathrm{K}$               & $300$ \\
        Neutral gas density          & $N_g$       & $\mathrm{m^{-3}}$          & $3.22 \times 10^{22}$ \\
        RF frequency                 & $f$         & $\mathrm{MHz}$             & $13.56$ \\
        RF voltage amplitude         & $V_0$       & $\mathrm{V}$               & $100$ \\
        \bottomrule
    \end{tabular}
\end{table}
The one-dimensional computational domain is discretized using a nonuniform symmetric mesh comprising $N_p=91$ nodes, yielding 90 cells. To adequately resolve the steep spatial gradients within the sheath regions, the node coordinates $x_i$ are densely clustered near both electrodes. Following the reference benchmark \cite{lymberopoulos1993fluid}, the grid distribution is mathematically defined as:

\begin{equation}
    \begin{gathered}
        x_i = \frac{L_x}{2} \left[ \frac{i-1}{(N_p-1)/2} \right]^2, \qquad i=1,\ldots,\frac{N_p-1}{2}+1, \\
        x_{N_p-i+1} = L_x-x_i, \qquad i=1,\ldots,\frac{N_p-1}{2}+1.
    \end{gathered}
\end{equation}

At the physical boundaries, a sinusoidal driving voltage is applied to the left powered electrode. For each discrete temporal collocation point $m$, the boundary potential is explicitly given by:
\begin{equation}
    \phi_{{\rm p},m} = V_0\sin(2\pi f t_m), 
\end{equation}
and the right grounded electrode remains fixed at $\phi_{\mathrm{g},m} = 0$. For the charged particle transport, the wall electron temperature is explicitly prescribed to evaluate the surface fluxes. The drift-diffusion transport of electrons and ions is governed by their respective mobilities and diffusivities, which are assumed to be inversely proportional to the neutral gas number density $N_g$.  The relevant transport coefficients, surface boundary parameters, and reaction thresholds are listed in Table~\ref{tab:transport_chemistry_params}.
\begin{table}[h]
    \centering
    \caption{Transport coefficients, surface boundary parameters, and reaction threshold energies.}
    \label{tab:transport_chemistry_params}
    \begin{tabular}{llll}
        \toprule
        \textbf{Parameter} & \textbf{Symbol} & \textbf{Unit} & \textbf{Value} \\
        \midrule
        Electron diffusivity product & $N_g D_e$   & $\mathrm{m^{-1}\,s^{-1}}$  & $3.86 \times 10^{24}$ \\
        Electron mobility product    & $N_g \mu_e$ & $\mathrm{V^{-1}\,m^{-1}\,s^{-1}}$    & $9.66 \times 10^{23}$ \\
        Ion diffusivity product      & $N_g D_i$   & $\mathrm{m^{-1}\,s^{-1}}$  & $2.07 \times 10^{20}$ \\
        Ion mobility product         & $N_g \mu_i$ & $\mathrm{V^{-1}\,m^{-1}\,s^{-1}}$    & $4.65 \times 10^{21}$ \\
        Wall electron temperature    & $T_{e,w}$   & $\mathrm{eV}$              & 0.5 \\
        Secondary emission coefficient& $\gamma$   & $-$                        & 0.01 \\
        Surface recombination velocity& $k_s$      & $\mathrm{m/s}$             & $1.19 \times 10^5$ \\
        Ionization threshold energy  & $\Delta\varepsilon_{\mathrm{ion}}$ & $\mathrm{eV}$ & $15.7$ \\
        Excitation threshold energy  & $\Delta\varepsilon_{\mathrm{exc}}$ & $\mathrm{eV}$ & $11.56$ \\
        \bottomrule
    \end{tabular}
\end{table}

Within the HB framework, the reaction rates are evaluated discretely and independently at each temporal collocation point $m$. Therefore, the reaction rates are expressed as:
\begin{align}
    R_{\mathrm{ion},m} &= N_g n_{e,m} k_{\mathrm{ion}}(\bar{\varepsilon}_{e,m}), \\
    R_{\mathrm{exc},m} &= N_g n_{e,m} k_{\mathrm{exc}}(\bar{\varepsilon}_{e,m}).
\end{align}

In these expressions, the local mean electron energy density at temporal collocation point $m$ is explicitly defined as:
\begin{equation}
    \bar{\varepsilon}_{e,m} = \frac{\varepsilon_{e,m}}{n_{e,m} e}.
\end{equation}

Utilizing this local mean energy, the corresponding rate coefficients $k_{\mathrm{ion}}$ and $k_{\mathrm{exc}}$ are logarithmically interpolated from the exact tabular data established in the reference benchmark \cite{lymberopoulos1993fluid}.

In the present study, two primary electron-impact reactions are considered. Consequently, direct ionization acts as the sole particle source:
\begin{equation}
    S_{e,m} = S_{i,m} = R_{\mathrm{ion},m},    
\end{equation}
while both reactions contribute to the inelastic electron energy loss. The total energy sink at each temporal collocation point is formulated as:
\begin{equation}
    S_{\varepsilon,m} = -\Delta\varepsilon_{\mathrm{ion}} R_{\mathrm{ion},m} -\Delta\varepsilon_{\mathrm{exc}} R_{\mathrm{exc},m},
\end{equation}
where the threshold energies $\Delta\varepsilon_{\mathrm{ion}}$ and $\Delta\varepsilon_{\mathrm{exc}}$ have been explicitly listed in Table~\ref{tab:transport_chemistry_params}.

\subsection{Establishment of the Time-Marching Reference Solution}
Before the dual-time stepping solution is employed as the reference for assessing the HB results, the convergence of the pseudo-time iterations must first be examined. Although the physical time step in the dual-time formulation is not restricted by the stability condition of equation system, an insufficiently converged inner solution introduces an additional iterative error into the physical-time update. The influence of the pseudo-CFL number and the required number of inner iterations is therefore evaluated at a representative physical time step $\Delta t=T/100$. To assess the convergence of the inner iterations, the normalized backward differentiation formula (BDF) residual, denoted as $\rho_s$, is monitored. This metric represents the $L_2$-norm of the complete residual vector---encompassing the electron number continuity, ion number continuity, and electron energy equations at the $s$-th pseudo-time iteration---normalized by its initial value at the start of the physical time step.

\begin{figure}[htbp]
    \centering
    \includegraphics[width=0.70\textwidth]{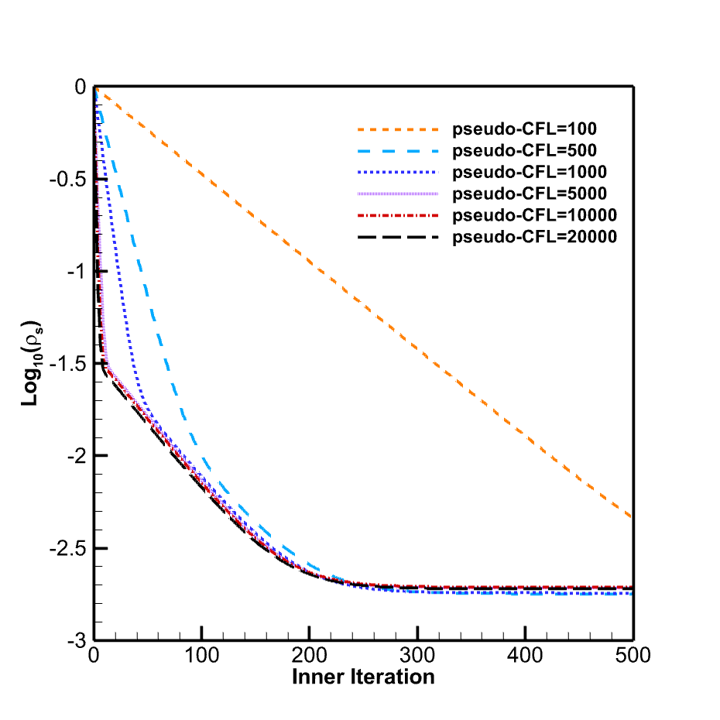}
    \caption{Convergence of the normalized full BDF residual at a representative physical time step for different pseudo-CFL numbers. The physical time step is $\Delta t=T/100$.}
    \label{fig:dt_inner_cfl}
\end{figure}

Figure~\ref{fig:dt_inner_cfl} shows that the convergence rate is strongly dependent on the pseudo-CFL number when a relatively small value is used. 

When the pseudo-CFL number is set to 100, the residual decreases slowly and fails to reach an asymptotic level even after 500 inner iterations. By raising this value to 500 and 1000, the initial reduction of the residual is substantially accelerated. Further increasing the pseudo-CFL number to the range of 5000--20000 yields a rapid drop of roughly 1.5 orders of magnitude within the first few iterations. After this initial sharp decrease, the solution enters a slower relaxation stage.

For pseudo-CFL numbers of 5000 and above, the convergence histories become virtually identical once the initial transient passes. In these cases, the residual settles at a plateau around $\log_{10}(\rho_s)\simeq-2.7$. The solver typically reaches this plateau within 250 to 300 inner iterations, and continuing the process up to 500 iterations yields no  improvement. Although the robust implicit formulation allows for a significantly higher stability limit, this consistent asymptotic behavior clearly demonstrates that pushing the pseudo-CFL number beyond 10000 offers no meaningful advantage for the current configuration. Guided by these findings, a pseudo-CFL number of 10000 and a maximum limit of 300 inner iterations per physical time step are selected. These parameters are adopted for the subsequent time-step verification and for the construction of the time-marching reference solution. 

With the pseudo-time parameters established, a physical-time-step refinement study is conducted to quantify the temporal discretization error. This step is essential to construct a sufficiently resolved time-marching baseline for evaluating the proposed time-domain HB method. Using the fixed pseudo-time configuration determined above, the temporal resolution is systematically refined by increasing the number of physical time steps per RF cycle as $T/\Delta t = 50,\ 100,\ 200,\ \text{and}\ 400$. Because the inner convergence is strictly controlled, the discrepancies among these test cases stem almost entirely from the physical-time integration, which is governed by the second-order backward differentiation formula.

Table~\ref{tab:dt_refinement} presents the spatially averaged electron number density, ion number density, and electron energy density sampled at the beginning of the converged RF cycle. To quantitatively assess the temporal discretization error, the calculation with the finest resolution ($T/\Delta t=400$) is adopted as the reference. The relative differences for each macroscopic quantity are evaluated as:
\begin{equation}
    E_{\mathrm{rel}}(q)
    =
    \frac{
    \left|\langle q\rangle_{\Delta t}
    -\langle q\rangle_{T/400}\right|
    }{
    \left|\langle q\rangle_{T/400}\right|
    }
    \times 100\%.
    \label{eq:dt_relative_error}
\end{equation}

\begin{table}[htbp]
    \centering
    \caption{Physical-time-step refinement based on the spatially
    averaged plasma quantities at $t/T=0$ of the converged RF period.
    Relative differences are calculated using the $T/\Delta t=400$
    solution as the reference.}
    \label{tab:dt_refinement}
    \begin{tabular}{ccccccc}
        \toprule
        $T/\Delta t$
        & $\langle n_e\rangle$
        & $E_{n_e}$
        & $\langle n_i\rangle$
        & $E_{n_i}$
        & $\langle\varepsilon_e\rangle$
        & $E_{\varepsilon_e}$ \\
        & $(10^{15}\,\mathrm{m}^{-3})$
        & $(\%)$
        & $(10^{15}\,\mathrm{m}^{-3})$
        & $(\%)$
        & $(10^{-3}\,\mathrm{J \cdot m}^{-3})$
        & $(\%)$ \\
        \midrule
        50  & 3.19145604 & 0.4798
            & 3.31838498 & 0.4634
            & 2.94977979 & 0.4624 \\
        100 & 3.17987346 & 0.1151
            & 3.30673529 & 0.1107
            & 2.93957294 & 0.1148 \\
        200 & 3.17696652 & 0.0236
            & 3.30382591 & 0.0226
            & 2.93689536 & 0.0236 \\
        400 & 3.17621689 & --    
            & 3.30307807 & --    
            & 2.93620186 & --     \\
        \bottomrule
    \end{tabular}
\end{table}

All three spatially averaged quantities converge monotonically as the physical time step is refined. Relative to the $T/\Delta t=400$ benchmark, the differences obtained with the coarsest resolution ($T/\Delta t=50$) are approximately $0.46$--$0.48\%$. These deviations rapidly decrease to approximately $0.11\%$ for $T/\Delta t=100$ and to less than $0.024\%$ for $T/\Delta t=200$. This uniform reduction in error across all three variables confirms that refining the time step smoothly and evenly improves the overall physical accuracy. To rigorously verify whether this consistent error reduction aligns with the theoretical expectations of the applied numerical approach, the observed temporal convergence order, $p$, is evaluated using three successively refined solutions:
\begin{equation}
    p
    =
    \log_2
    \left[
    \frac{
    \left|\langle q\rangle_{N}
    -\langle q\rangle_{2N}\right|
    }{
    \left|\langle q\rangle_{2N}
    -\langle q\rangle_{4N}\right|
    }
    \right],
    \label{eq:observed_temporal_order}
\end{equation}
where $N=T/\Delta t$ represents the temporal resolution per RF cycle. Evaluated with the coarser triplet ($N=50$, 100, and 200), the observed orders for $n_e$, $n_i$, and $\varepsilon_e$ are calculated as $1.994$, $2.002$, and $1.931$, respectively. When utilizing the finer triplet ($N=100$, 200, and 400), these estimates become $1.955$, $1.960$, and $1.949$. All computed values exhibit excellent agreement with the theoretical second-order accuracy of the used temporal discretization.

\begin{figure}[h]
    \centering
    \includegraphics[width=0.70\textwidth]
    {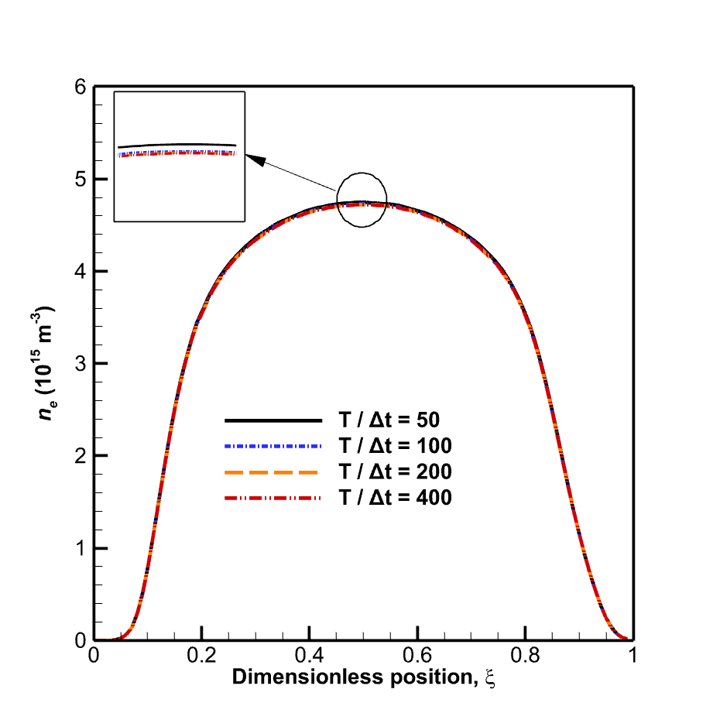}
    \caption{Electron number density distributions at the beginning
    of the converged RF period for different physical-time-step
    resolutions.}
    \label{fig:dt_spatial_refinement}
\end{figure}

To offer a visual assessment of the time-step refinement, Fig.~\ref{fig:dt_spatial_refinement} compares the spatial distributions of the electron number density sampled at the beginning of the converged RF cycle.
All four solutions yield nearly identical sheath-edge locations near the electrodes. As highlighted by the inset, the remaining temporal discretization error is confined primarily to the bulk-density plateau. The coarsest resolution ($T/\Delta t=50$) slightly overpredicts the bulk density, whereas the profiles for $T/\Delta t=100$, 200, and 400 progressively converge. Notably, the $T/\Delta t=200$ and 400 curves are almost indistinguishable across the entire discharge gap, perfectly aligning with the negligible relative difference (under $0.024\%$) reported in Table~\ref{tab:dt_refinement}. Moreover, relaxing the resolution from 400 to 200 steps per cycle cuts the dominant time-marching cost roughly in half, assuming the same number of inner iterations.

Accordingly, the configuration utilizing $T/\Delta t=200$, a pseudo-CFL number of 10000, and $n_{\mathrm{inner}}=300$ is selected as the reference DTS baseline. This setup will be used to evaluate both the accuracy and the computational efficiency of the proposed time-domain HB method.

\subsection{A Priori Spectral Analysis and Harmonic Truncation}

Before performing the HB calculations, an \textit{a priori} spectral analysis is conducted to determine the optimal number of harmonics required to accurately resolve the RF CCP dynamics. This preliminary analysis relies on the high-fidelity transient signals extracted directly from the reference DTS baseline established in the preceding section. 

\begin{figure}[htbp]
    \centering
    \begin{minipage}{0.48\textwidth}
        \centering
        \includegraphics[width=\textwidth]{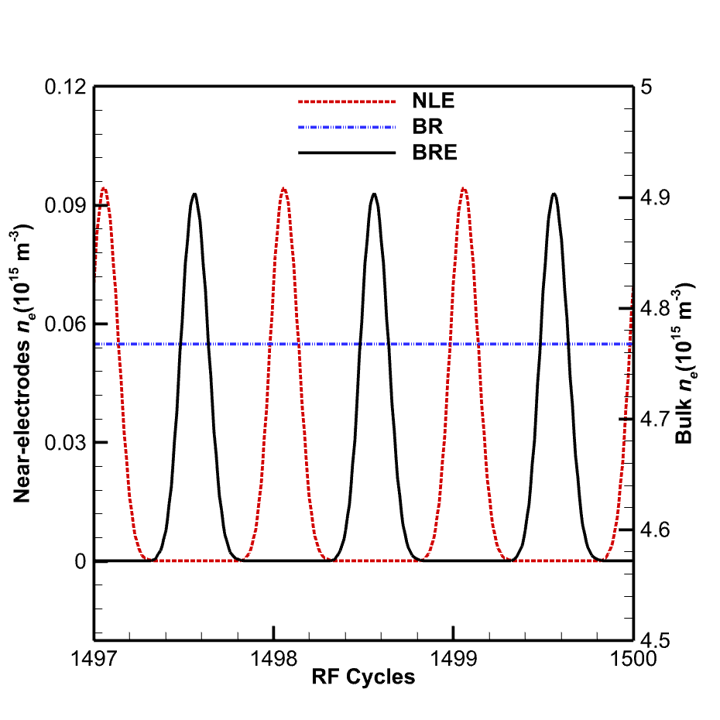}
        \caption{Time-domain signals of electron number density at different monitoring points.}
        \label{fig:td_signals_ne}
    \end{minipage}\
    \begin{minipage}{0.48\textwidth}
        \centering
        \includegraphics[width=\textwidth]{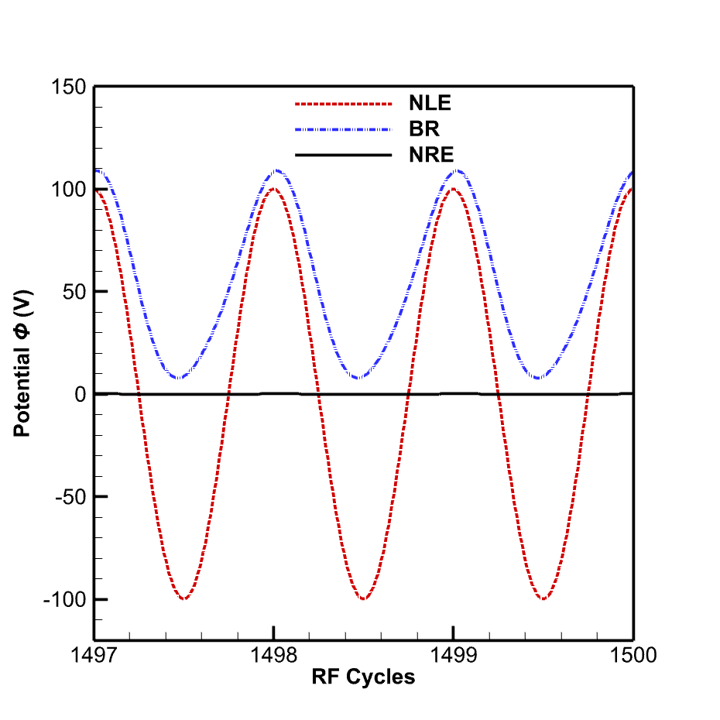}
        \caption{Time-domain signals of electric potential at different monitoring points.}
        \label{fig:td_signals_phi}
    \end{minipage}
\end{figure}

To capture the spatiotemporal evolution of the discharge, three representative spatial locations are monitored: the near-left electrode (NLE), the bulk region (BR), and the near-right electrode (NRE). The temporal variations of the electron number density and the electric potential over consecutive RF cycles are illustrated in Fig.~\ref{fig:td_signals_ne} and Fig.~\ref{fig:td_signals_phi}, respectively.

The time-domain signals reveal that the $n_e$ waveforms at both the NLE and NRE exhibit distinct periodic oscillations. Because these profiles deviate significantly from pure sinusoids, they naturally encompass higher-order harmonic components. Although they experience a phase shift in the time domain, their frequency-domain amplitude spectra remain identical. In contrast, the electron density in the BR stays nearly constant; its signal is dominated almost entirely by the time-averaged component, leaving the harmonic amplitudes virtually at zero. Similarly, because the NRE is strictly grounded ($\phi = 0$), its electric potential inherently carries zero harmonic content. Consequently, the subsequent Fast Fourier Transform (FFT) analysis focuses selectively on the NLE for $n_e$, and on both the NLE and BR for $\phi$.

\begin{figure}[h]
    \centering
    \includegraphics[width=0.7\textwidth]{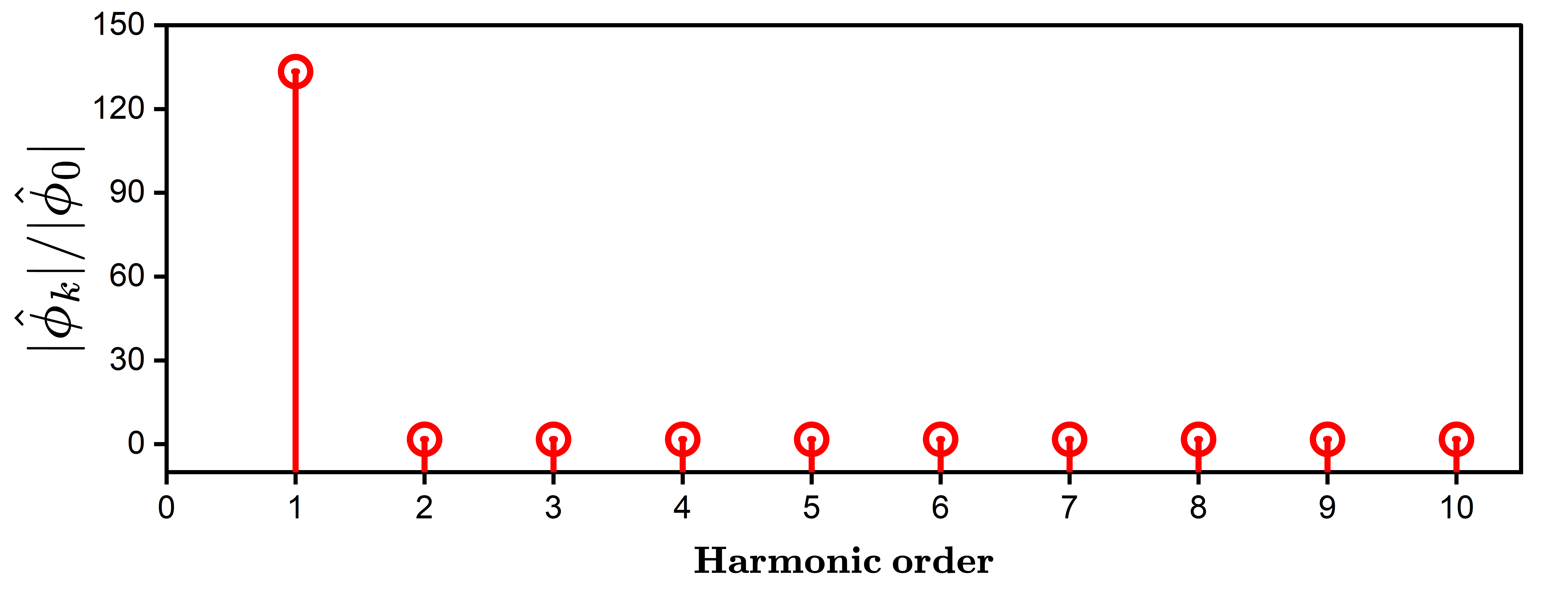}
    \caption{Normalized FFT spectrum of potential at NLE.}
    \label{fig:fft_phi_nle}
\end{figure}

\begin{figure}[h]
    \centering
    \includegraphics[width=0.7\textwidth]{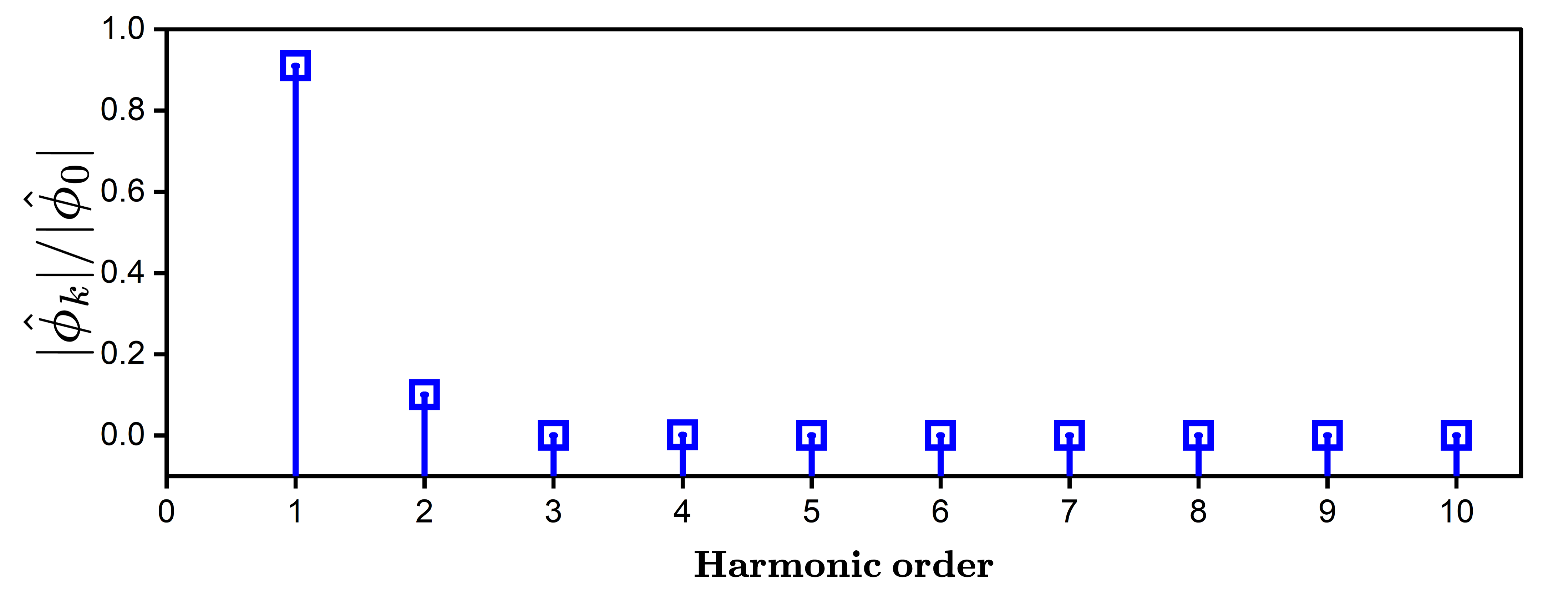}
    \caption{Normalized FFT spectrum of potential at BR.}
    \label{fig:fft_phi_br}
\end{figure}

\begin{figure}[h]
    \centering
    \includegraphics[width=0.7\textwidth]{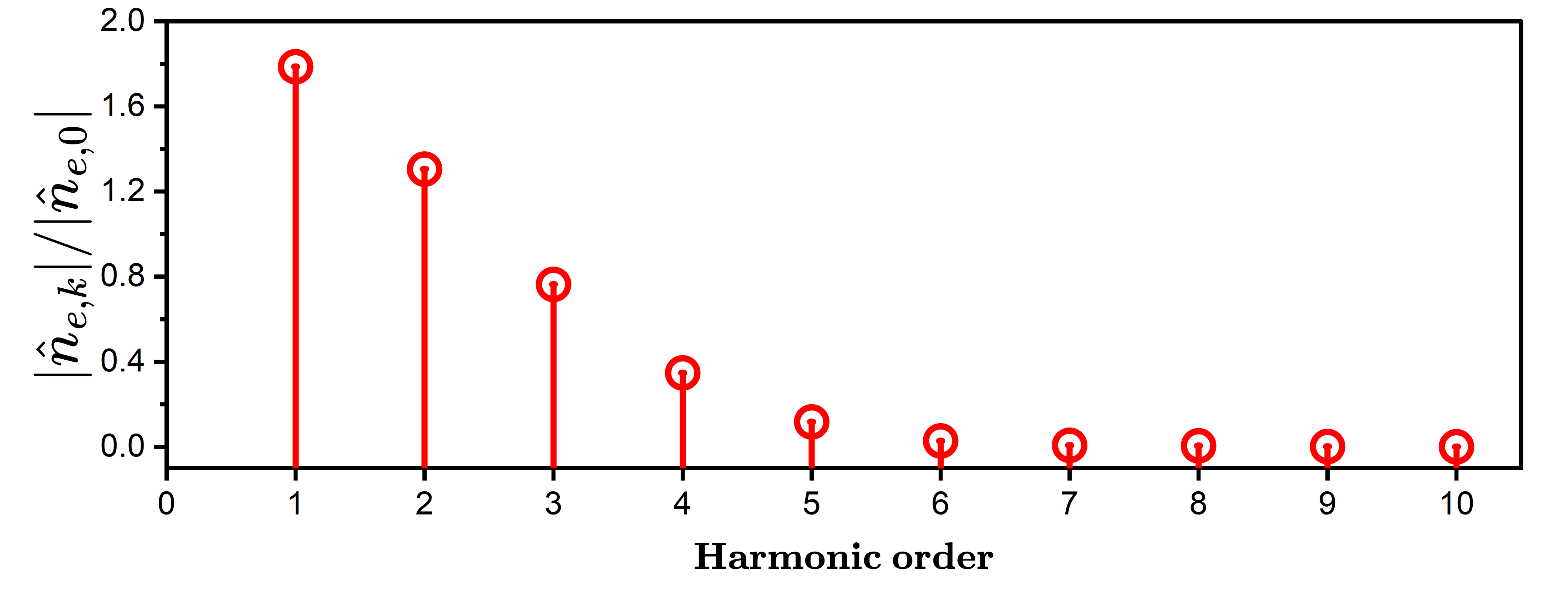}
    \caption{Normalized FFT spectrum of electron number density at NLE.}
    \label{fig:fft_ne_nle}
\end{figure}

The resulting spectral amplitude distributions, normalized by their respective time-averaged values, are presented in Figs.~\ref{fig:fft_phi_nle}--\ref{fig:fft_ne_nle}. The electric potential is composed primarily of low-order harmonics. Specifically, the NLE potential is almost exclusively dominated by the fundamental frequency ($N_H=1$). The BR potential, however, is influenced by the nonlinear response of the bulk plasma, necessitating the inclusion of both the first and second harmonics ($N_H=2$). Conversely, driven by the intense dynamics of sheath depletion and expansion, the electron number density at the NLE exhibits a much broader spectral distribution. The FFT spectrum of $n_e$ demonstrates that the normalized amplitudes decay progressively across the first five harmonics, diminishing to negligible levels from the sixth harmonic onward.

\begin{figure}[h]
    \centering
    \begin{minipage}{0.48\textwidth}
        \centering
        \includegraphics[width=\textwidth]{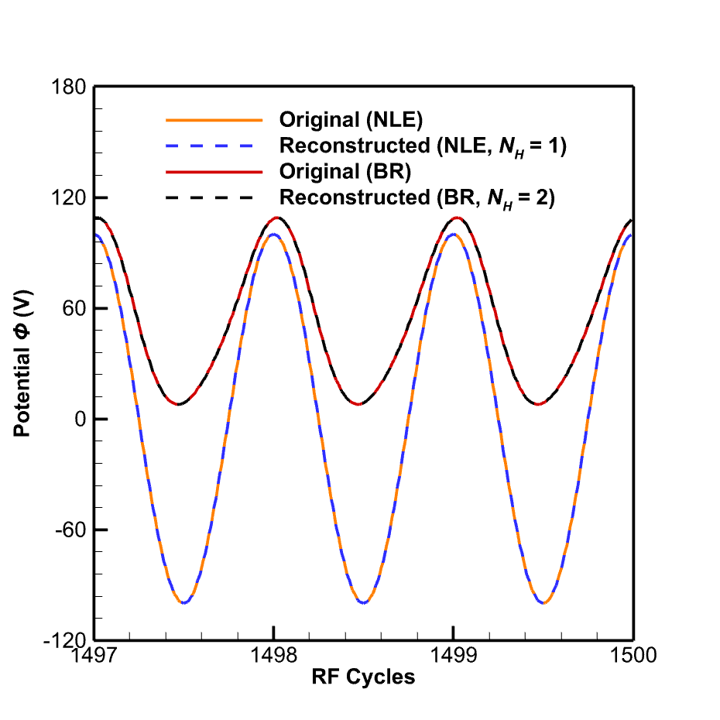}
        \caption{Truncated reconstruction of potential using dominant harmonics.}
        \label{fig:ori_vs_recon_phi}
    \end{minipage}\hfill
    \begin{minipage}{0.48\textwidth}
        \centering
        \includegraphics[width=\textwidth]{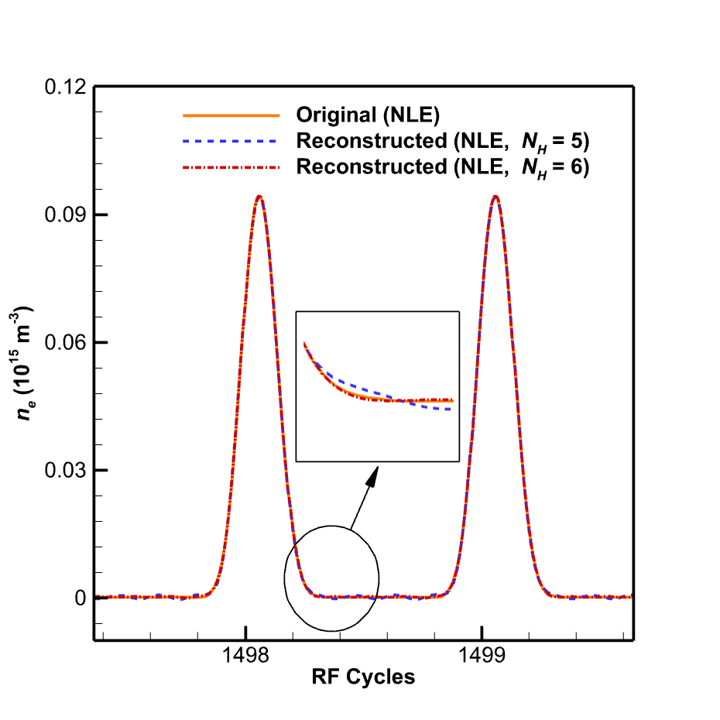}
        \caption{Truncated reconstruction of NLE electron number density using $N_H=5$.}
        \label{fig:ori_vs_recon_ne}
    \end{minipage}
\end{figure}

To quantitatively verify this frequency truncation, inverse FFT reconstructions are performed using the selected dominant harmonics (Figs.~\ref{fig:ori_vs_recon_phi} and \ref{fig:ori_vs_recon_ne}). As expected, truncating at $N_H=1$ for the NLE potential and $N_H=2$ for the BR potential demonstrates excellent agreement with the original time-accurate curves. For the electron number density at the NLE, a reconstruction utilizing the first five harmonics ($N_H=5$) successfully captures the overall shape and magnitude of the original signal. However, due to the steep temporal gradients within the sheath, localized spurious oscillations remain visible near the density minima. Increasing the number of harmonics to $N_H=6$ effectively suppresses these spurious oscillations, yielding a smooth and faithful reconstruction. Guided by this \textit{a priori} spectral analysis, the subsequent implicit time-domain HB computations are systematically evaluated across four harmonic truncation levels: $N_H = 6, 8, 10$, and $12$.

\subsection{Harmonic Convergence and Accuracy Assessment}

To evaluate the numerical performance of the proposed method, the implicit HB computations are performed across four selected harmonic truncation levels ($N_H = 6, 8, 10$, and $12$). Preliminary computations revealed that employing lower harmonic counts, such as $N_H = 4$ or $5$, inevitably leads to numerical divergence. This instability is primarily triggered by unresolved nonlinear aliasing and spurious numerical oscillations within the highly depleted sheath regions, which quickly destabilize the tightly coupled governing equation system. Therefore, $N_H=6$ establishes the minimum threshold for a stable numerical integration. In the time-domain HB formulation, these selected harmonic counts correspond to resolving the plasma dynamics at $N_T = 2N_H + 1$ discrete temporal collocation points (i.e., $N_T = 13, 17, 21$, and $25$, respectively) per RF cycle.

\begin{figure}[h]
    \centering
    \includegraphics[width=0.7\textwidth]{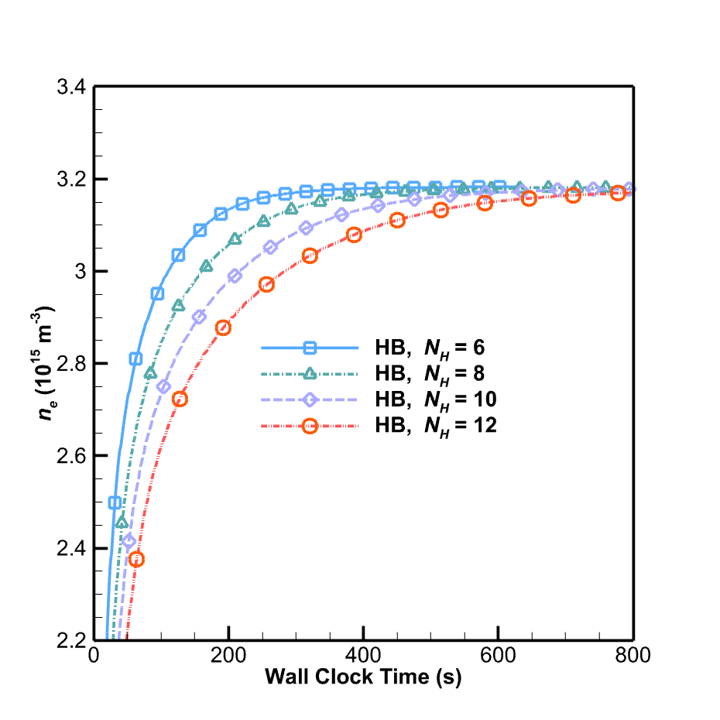}
    \caption{Convergence histories of the spatially averaged electron number density at the beginning of the RF cycle for different harmonic truncation levels.}
    \label{fig:harmonic_convergence}
\end{figure}

Benefiting from the robust implicit relaxation approach, the pseudo-time marching circumvents strict stability restrictions, permitting the use of an aggressive pseudo-CFL number of $10000$ (consistent with the previously established baseline). Figure~\ref{fig:harmonic_convergence} illustrates the convergence histories of the spatially averaged electron number density at the beginning of the RF cycle as a function of the physical wall-clock time.

As depicted, the macroscopic plasma state rapidly converges to a stable periodic solution. Regardless of the harmonic count, all cases reach an identical converged value of approximately $3.179 \times 10^{15}~\mathrm{m^{-3}}$. This rigorous consistency corroborates the preceding spectral analysis, confirming that retaining only the primary harmonics is  sufficient to yield an accurate spatially averaged solution.

\begin{figure}[!h]
    \centering
    \begin{subfigure}[b]{0.48\textwidth}
        \centering
        \includegraphics[width=\textwidth]{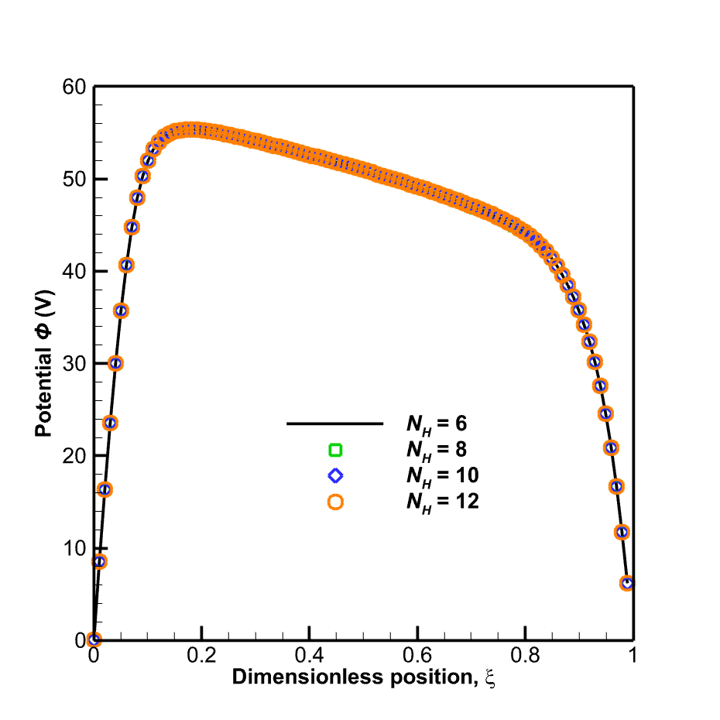}
        \caption{Electric potential ($\phi$)}
        \label{subfig:phi_NH}
    \end{subfigure}\hfill
    \begin{subfigure}[b]{0.48\textwidth}
        \centering
        \includegraphics[width=\textwidth]{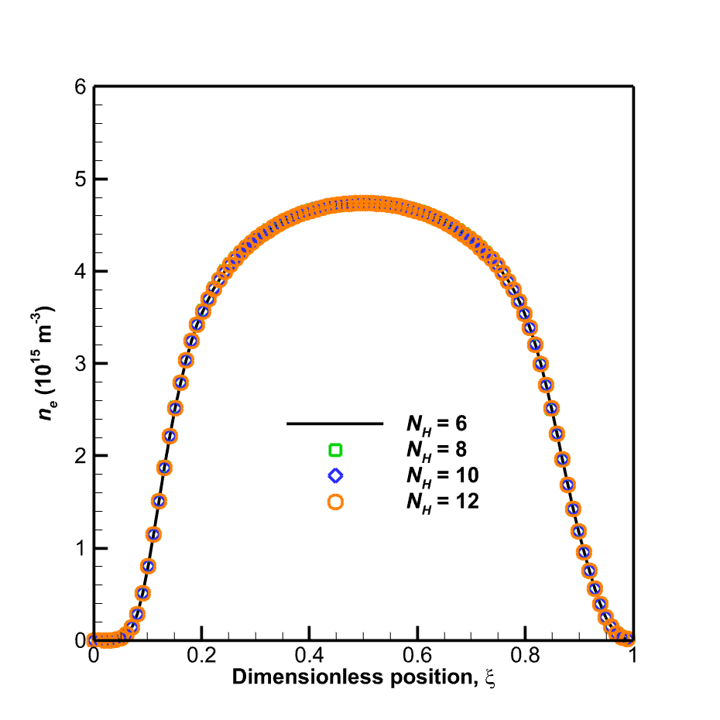}
        \caption{Electron density ($n_e$)}
        \label{subfig:ne_NH}
    \end{subfigure}
    
    \begin{subfigure}[b]{0.48\textwidth}
        \centering
        \includegraphics[width=\textwidth]{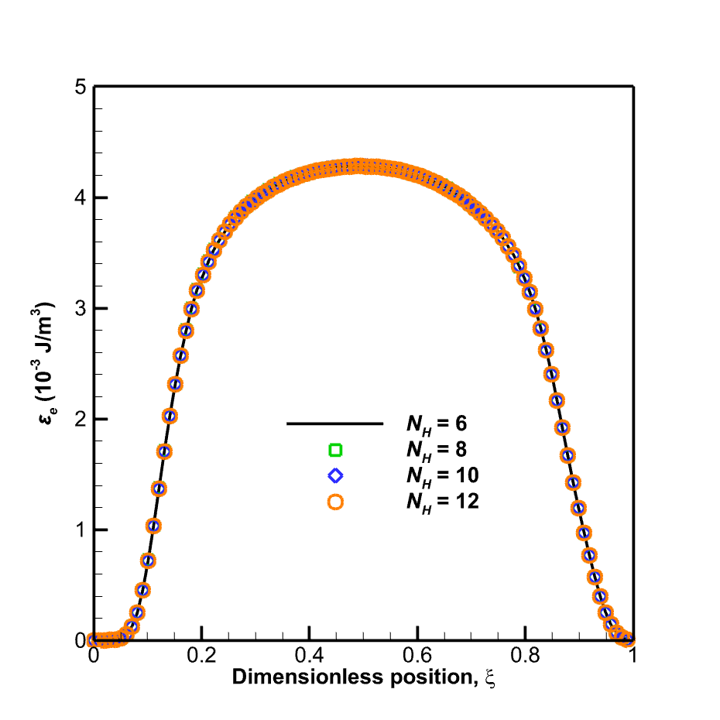}
        \caption{Electron energy density ($\varepsilon_e$)}
        \label{subfig:Ee_NH}
    \end{subfigure}\hfill
    \begin{subfigure}[b]{0.48\textwidth}
        \centering
        \includegraphics[width=\textwidth]{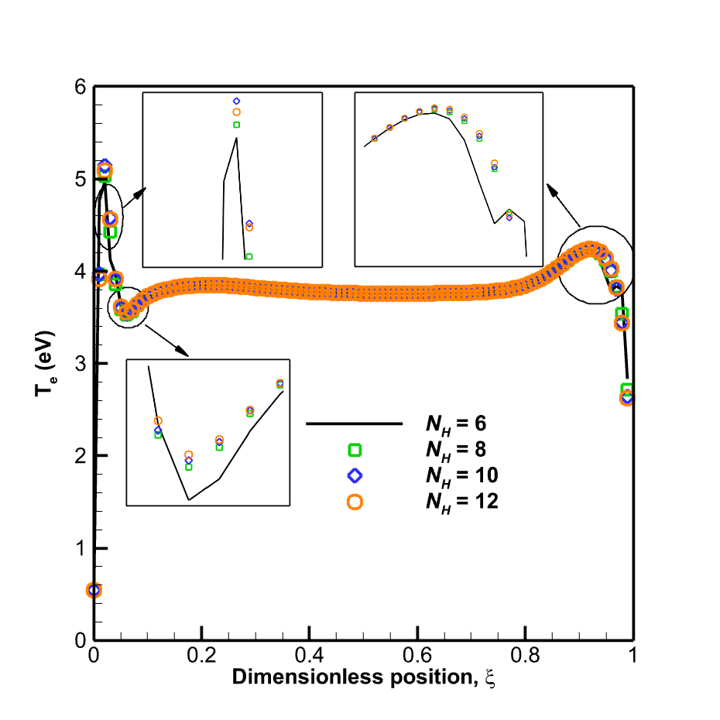}
        \caption{Electron temperature ($T_e$)}
        \label{subfig:Te_NH}
    \end{subfigure}
    \caption{Spatial distributions of macroscopic plasma properties at the beginning of the RF cycle, reconstructed using $N_H = 6, 8, 10$, and $12$.}
    \label{fig:spatial_convergence_profiles}
\end{figure}

To further assess the solution accuracy, the reconstructed profiles of the electric potential, electron number density, electron energy density, and electron temperature sampled at the beginning of the RF cycle are compared in Fig.~\ref{fig:spatial_convergence_profiles}.

As shown in Figs.~\ref{subfig:phi_NH}--\ref{subfig:Ee_NH}, the spatial profiles of the electric potential, electron number density, and electron energy density exhibit excellent agreement across all tested harmonic counts. This consistent behavior demonstrates that the primary transport mechanisms and the overall discharge structure are well captured, even with the lowest selected harmonic count ($N_H=6$).

However, a minor deviation is visible in the electron temperature profile for $N_H = 6$ within the near-wall sheath regions (Fig.~\ref{subfig:Te_NH}). Because $T_e$ is a derived property computed from the ratio of energy density to particle number density, its calculation is inherently sensitive to minor numerical perturbations in these highly depleted zones. Although the underlying conservative variables ($n_e$ and $\varepsilon_e$) remain robust and largely unaffected, a slightly higher harmonic count is preferred to guarantee smooth and artifact-free derived properties. 

This requirement for higher accuracy must be carefully weighed against the increased computational cost. Since increasing the harmonic count enlarges the coupled HB system, the computational cost grows accordingly. To provide a fair timing assessment, these tests were executed on a single core of an Intel Core Ultra 9 285H processor. The $N_H = 6$ case achieves full convergence in approximately $600$~s, whereas the highest-resolution case ($N_H = 12$) doubles this duration to roughly $1200$~s. To eliminate the minor temperature deviations without incurring the severe computational penalty of higher-order truncations, $N_H = 8$ is selected. This configuration strikes an optimal balance between accuracy and efficiency, serving as the verified baseline for the subsequent spatiotemporal analyses.

\subsection{Spatiotemporal Validation of the Time-Domain Harmonic Balance Method}

\begin{figure}[h]
    \centering
    \begin{subfigure}[b]{0.5\textwidth}
        \centering
        \includegraphics[width=\textwidth]{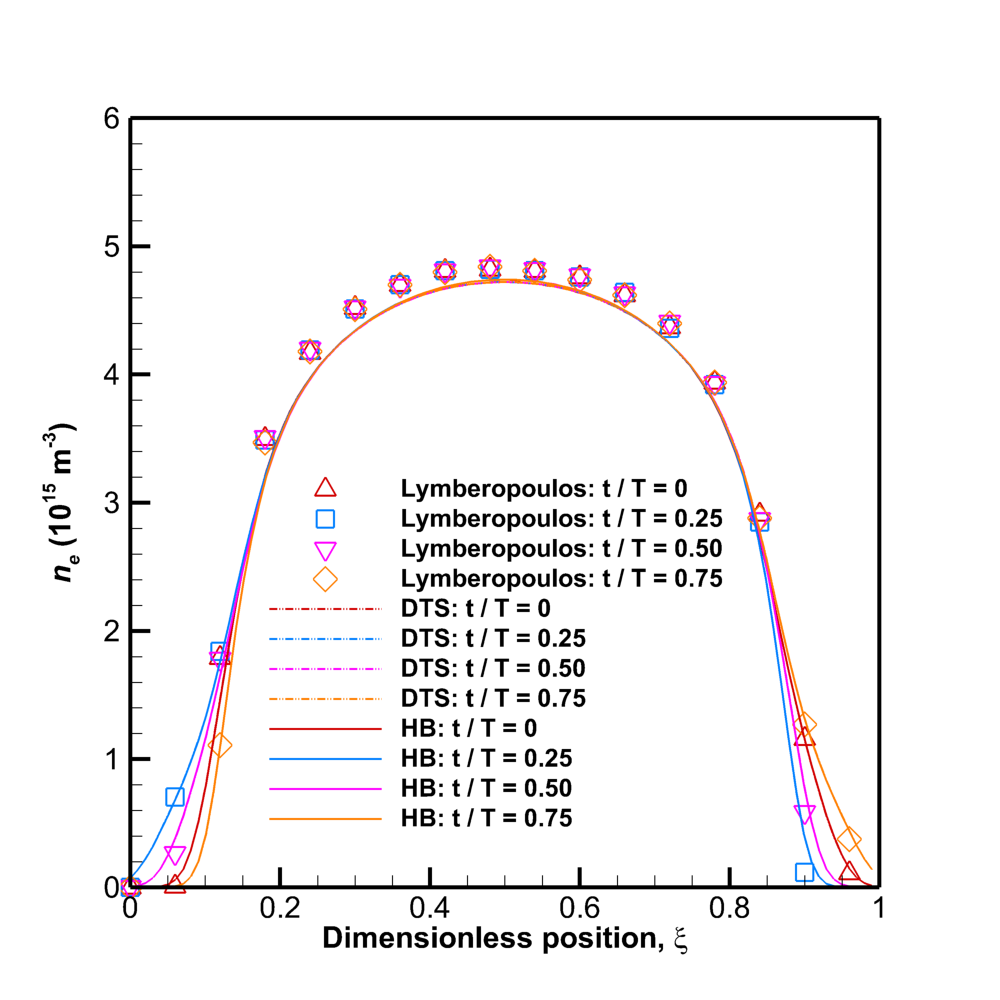}
        \caption{Electron density ($n_e$)}
        \label{subfig:ne_DTS_HB}
    \end{subfigure}\hfill
    \begin{subfigure}[b]{0.5\textwidth}
        \centering
        \includegraphics[width=\textwidth]{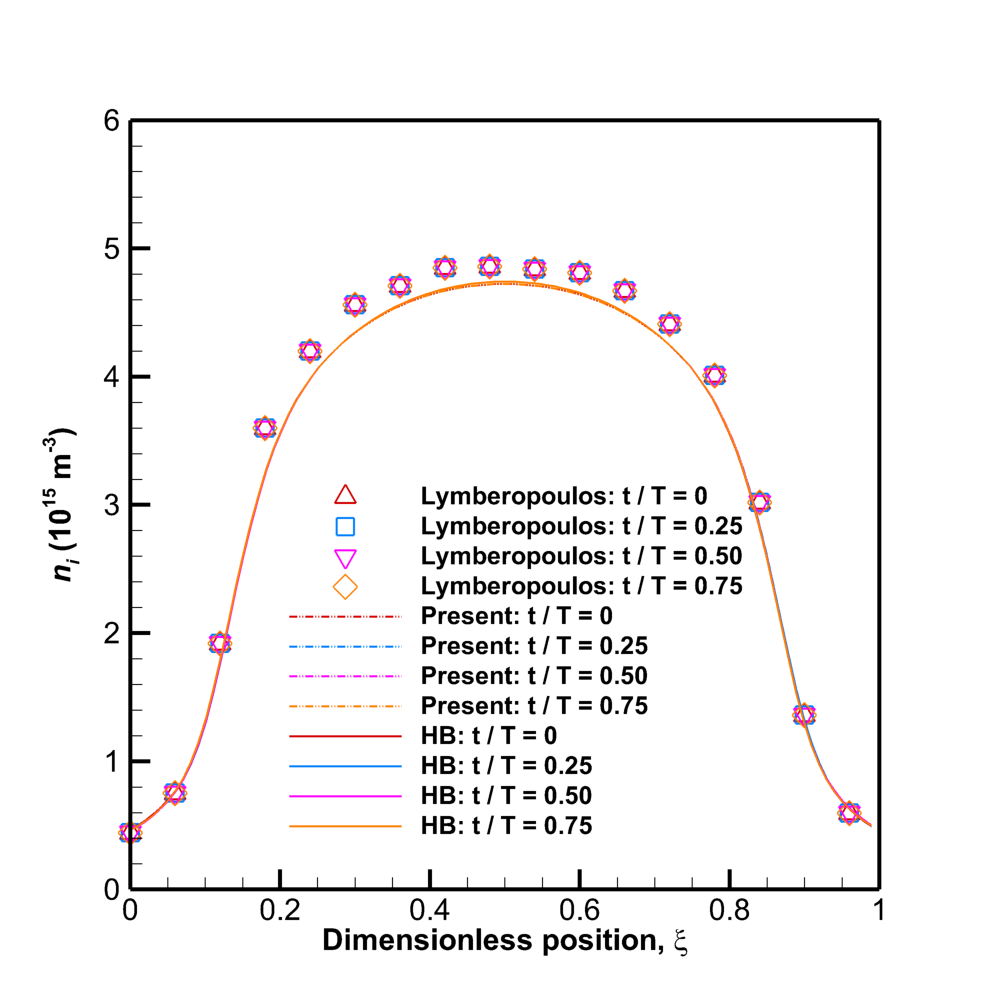}
        \caption{Ion density ($n_i$)}
        \label{subfig:ni_DTS_HB}
    \end{subfigure}
    \caption{Comparison of spatial distributions of (a) electron number density and (b) ion number density at different phases of the RF cycle among the 1993 reference data, the DTS method, and the proposed  time-domain HB method ($N_H=8$).}
    \label{fig:spatial_profiles_comparison}
\end{figure}

To rigorously validate the proposed time-domain HB method, the spatiotemporal evolutions of the charged particle densities are compared against both the conventional DTS solutions and the widely recognized benchmark data of Lymberopoulos and Economou \cite{lymberopoulos1993fluid}. To ensure a strict and fair comparison, the DTS results utilized here are generated using the rigorously verified baseline configuration established in the preceding section ($T/\Delta t = 200$ and $n_{\mathrm{inner}} = 300$). Figure~\ref{fig:spatial_profiles_comparison} illustrates the spatial distributions of the electron and ion number densities sampled at four equidistant phases of the RF cycle: $t/T = 0, 0.25, 0.50$, and $0.75$.

As shown in Fig.~\ref{fig:spatial_profiles_comparison}, the time-domain HB method captures the dynamic expansion and contraction of the sheath regions with high fidelity, demonstrating excellent agreement with both the time-domain DTS solutions and the benchmark data. In the bulk plasma region, the peak particle number densities computed by the DTS baseline and the proposed HB method exhibit perfect mutual consistency. Both approaches align closely with the 1993 reference, yielding a minor relative deviation of within $2\%$.

This successful validation against the reference data~\cite{lymberopoulos1993fluid} confirms the physical configuration and the overall macroscopic discharge structure. Any minor remaining discrepancies can be reasonably attributed to external factors, such as differences in the spatial discretization schemes, the interpolation of tabulated reaction rates, and temporal integration details absent from the original study. Crucially, these external sources of discrepancy must be clearly distinguished from the internal consistency of the proposed methods; both the HB and DTS solvers employ an identical computational mesh, finite-volume method, chemistry model, and boundary conditions.

\begin{figure*}[htbp]
    \centering
    \begin{subfigure}[t]{0.48\textwidth}
        \centering
        \includegraphics[width=\textwidth]
        {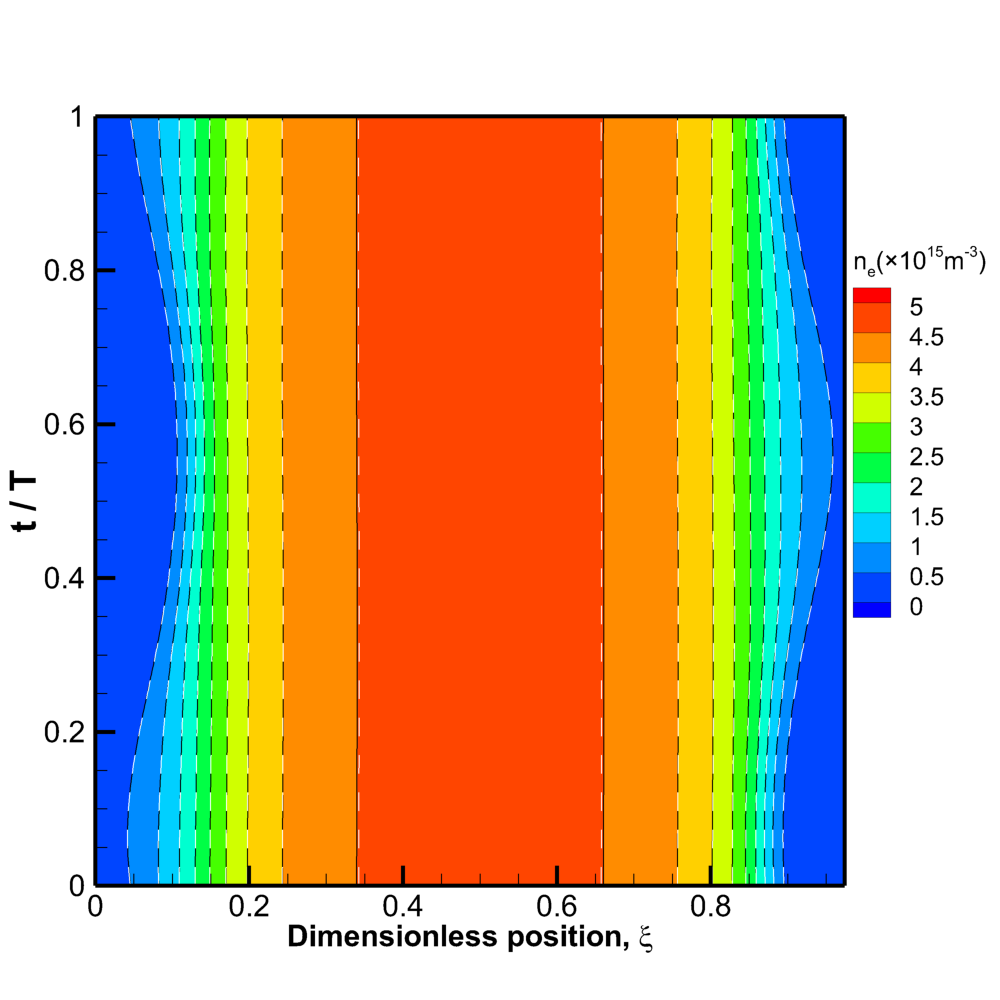}
        \caption{Electron number density, $n_e$.}
        \label{fig:spacetime_ne}
    \end{subfigure}
    \hfill
    \begin{subfigure}[t]{0.48\textwidth}
        \centering
        \includegraphics[width=\textwidth]
        {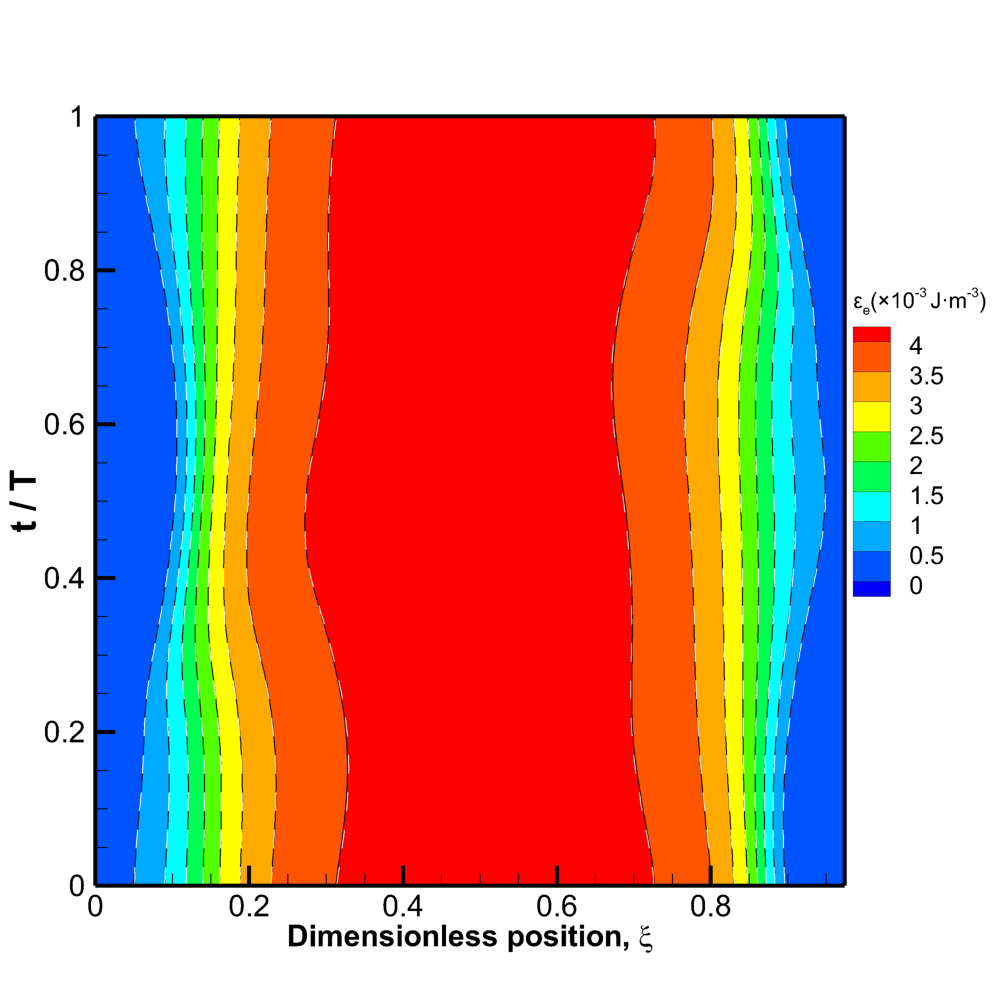}
        \caption{Electron energy density, $\varepsilon_e$.}
        \label{fig:spacetime_ee}
    \end{subfigure}

    \begin{subfigure}[t]{0.48\textwidth}
        \centering
        \includegraphics[width=\textwidth]
        {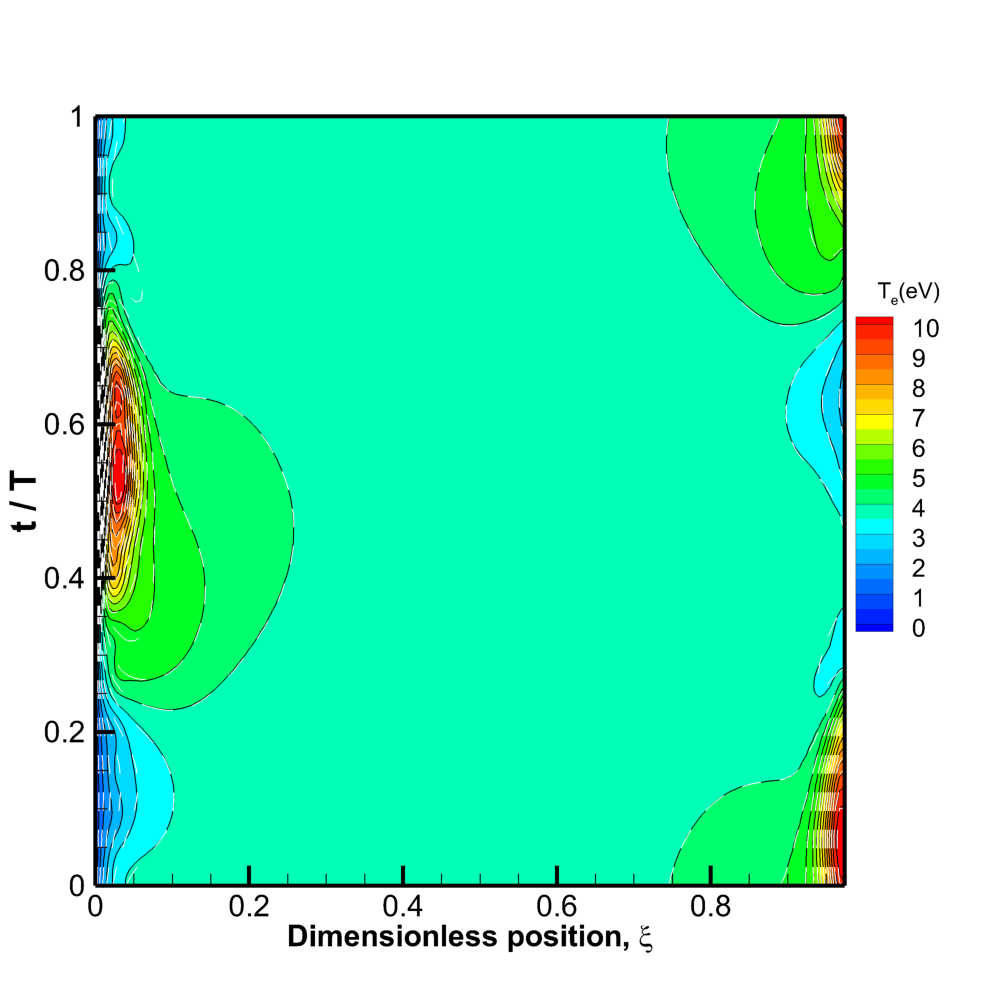}
        \caption{Electron temperature, $T_e$.}
        \label{fig:spacetime_te}
    \end{subfigure}
    \hfill
    \begin{subfigure}[t]{0.48\textwidth}
        \centering
        \includegraphics[width=\textwidth]
        {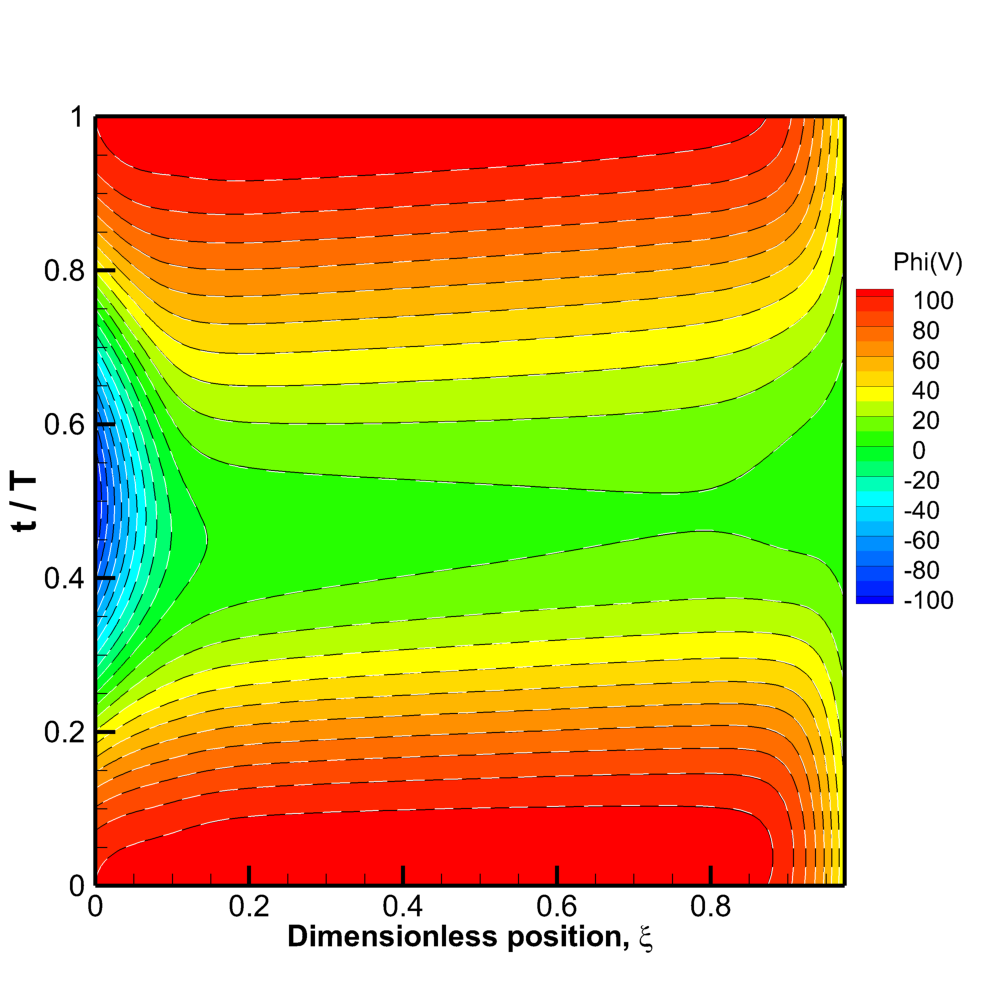}
        \caption{Electric potential, $\phi$.}
        \label{fig:spacetime_phi}
    \end{subfigure}

    \caption{Spatiotemporal comparison between the DTS reference
    solution and the HB solution over one RF period. Filled contours
    represent the DTS solution with $T/\Delta t=200$, while the white
    isolines represent the reconstructed HB solution with $N_H=8$.
    The spatial and temporal coordinates are normalized as
    $\xi=x/L$ and $t/T$, respectively.}
    \label{fig:spacetime_comparison}
\end{figure*}

However, demonstrating agreement at four isolated phases is insufficient to confirm whether the complete periodic response is accurately reproduced. To address this, the spatiotemporal evolutions computed by the time-domain HB and DTS methods are compared comprehensively throughout the entire RF cycle in Fig.~\ref{fig:spacetime_comparison}. The filled contours represent the fully converged DTS baseline (utilizing $T/\Delta t=200$), while the overlaid white isolines denote the HB solution reconstructed using $N_H=8$. To ensure a direct and rigorous evaluation, identical contour levels are applied to both solutions.

The spatiotemporal evolution of the electron number density (Fig.~\ref{fig:spacetime_ne}) reveals a relatively stationary, high-density bulk plasma bounded by periodically modulated sheaths. The dynamic displacement of the sheath edges is clearly reflected in the pronounced curvature of the near-electrode contours. Across the entire RF cycle, the HB isolines perfectly track the underlying DTS reference, even within zones of steep spatiotemporal gradients, confirming that the truncated harmonic set resolves both the quasi-steady bulk and the highly nonlinear transient sheath dynamics.

Broadly correlated with $n_e$, the electron energy density (Fig.~\ref{fig:spacetime_ee}) nonetheless exhibits a more pronounced temporal asymmetry. This distinction arises from the complex interplay among field-driven heating, energy transport, and collisional losses. The proposed solver faithfully reproduces the high-energy bulk and the periodic contour deformations near the electrodes, demonstrating that the HB formulation's high fidelity naturally extends to the full electron-energy balance.

By contrast, the electron temperature (Fig.~\ref{fig:spacetime_te}) displays distinct characteristics stemming directly from its nature as a derived ratio ($T_e \propto \varepsilon_e/n_e$). While the bulk plasma maintains a narrow temperature band, localized heating maxima alternately emerge near the electrodes during intense sheath excitation. The HB method accurately captures the location, phase, and extent of these heating zones. However, as anticipated, the localized deviations between the two solutions are slightly magnified in the strongly depleted sheaths, where minute numerical differences in the conservative variables are amplified by the division operation. Consequently, the near-wall $T_e$ profiles inherently warrant a more cautious interpretation than the directly integrated fields.

Finally, the electric potential (Fig.~\ref{fig:spacetime_phi}) is strongly modulated by the applied RF voltage, featuring smooth bulk variations but dynamically emerging steep gradients as the voltage drop alternately redistributes between the sheaths. The excellent alignment between the overlaid isolines and the reference contours confirms that the proposed formulation accurately captures the phase and amplitude of this electrostatic response. Accurately resolving $\phi$ is of paramount importance; as it governs the drift fluxes and heating terms, the potential acts as the primary driver coupling the macroscopic plasma state across all discrete temporal collocation points.

To complement the visual comparisons with a rigorous quantitative assessment, the volume-weighted spatiotemporal relative $L_2$ error is defined as:
\begin{equation}
    E_2(q)=
    \left[
    \frac{
    \displaystyle
    \sum_{m=0}^{N_s-1}\sum_{i=1}^{N_c}
    V_i\left(q_{i,m}^{\mathrm{HB}}-
    q_{i,m}^{\mathrm{DTS}}\right)^2
    }{
    \displaystyle
    \sum_{m=0}^{N_s-1}\sum_{i=1}^{N_c}
    V_i\left(q_{i,m}^{\mathrm{DTS}}\right)^2
    }
    \right]^{1/2},
    \label{eq:spacetime_l2}
\end{equation}
where $N_s=200$ represents the number of distinct temporal phase samples per RF cycle. The normalized mean absolute error ($E_{\mathrm{MA}}$) is defined analogously as:
\begin{equation}
    E_{\mathrm{MA}}(q)=
    \frac{
    \displaystyle
    \sum_{m=0}^{N_s-1}\sum_{i=1}^{N_c}
    V_i\left|q_{i,m}^{\mathrm{HB}}-
    q_{i,m}^{\mathrm{DTS}}\right|
    }{
    \displaystyle
    \sum_{m=0}^{N_s-1}\sum_{i=1}^{N_c}
    V_i\left|q_{i,m}^{\mathrm{DTS}}\right|
    }.
    \label{eq:spacetime_nmae}
\end{equation}

Additionally, the maximum pointwise relative error across the entire spatiotemporal domain is evaluated and denoted as $E_{\infty}(q)$.

\begin{table}[htbp]
    \centering
    \caption{Quantitative spatiotemporal error metrics of the HB solution ($N_H=8$) evaluated against the fully converged DTS baseline ($T/\Delta t=200$).}
    \label{tab:spacetime_errors}
    \begin{tabular}{lccc}
        \toprule
        Variable
        & $E_2$ (\%)
        & $E_{\mathrm{MA}}$ (\%)
        & $E_{\infty}$ (\%) \\
        \midrule
        $n_e$             & 0.258 & 0.222 & 0.392 \\
        $\varepsilon_e$   & 0.236 & 0.208 & 0.371 \\
        $T_e$             & 1.180 & 0.334 & 6.385 \\
        $\phi$            & 0.167 & 0.175 & 0.227 \\
        \bottomrule
    \end{tabular}
\end{table}

As summarized in Table~\ref{tab:spacetime_errors}, the relative $L_2$ and mean absolute errors for the directly integrated variables ($n_e$, $\varepsilon_e$, and $\phi$) are remarkably small, all remaining strictly below $0.3\%$. While the maximum pointwise error ($E_\infty$) for the electron temperature reaches approximately $6.4\%$, this extreme deviation is strictly confined to the near-wall cells. As previously discussed, this localized peak error is simply a mathematical consequence of dividing two extremely small quantities ($\varepsilon_e$ and $n_e$) to calculate $T_e$. It does not mean the underlying electron-energy calculation itself is inaccurate. Taken together, the isolated phase-resolved profiles, the complete spatiotemporal distributions, and these comprehensive error metrics conclusively demonstrate that retaining $N_H=8$ harmonics is perfectly sufficient to accurately reproduce the rigorous periodic DTS solution for the present RF discharge.

\subsection{Computational Efficiency Analyses}

Unlike conventional physical-time integration, the HB formulation inherently bypasses physical time-step errors, introducing instead harmonic-truncation and nonlinear-aliasing errors. As evidenced by the preceding spatiotemporal validation, these spectral errors remain negligible for $N_H=8$. Crucially, this high spectral accuracy is achieved while entirely circumventing the exhaustive physical transients required to reach a time-asymptotic periodic state.

\begin{figure}[!h]
    \centering
    \includegraphics[width=0.7\textwidth]{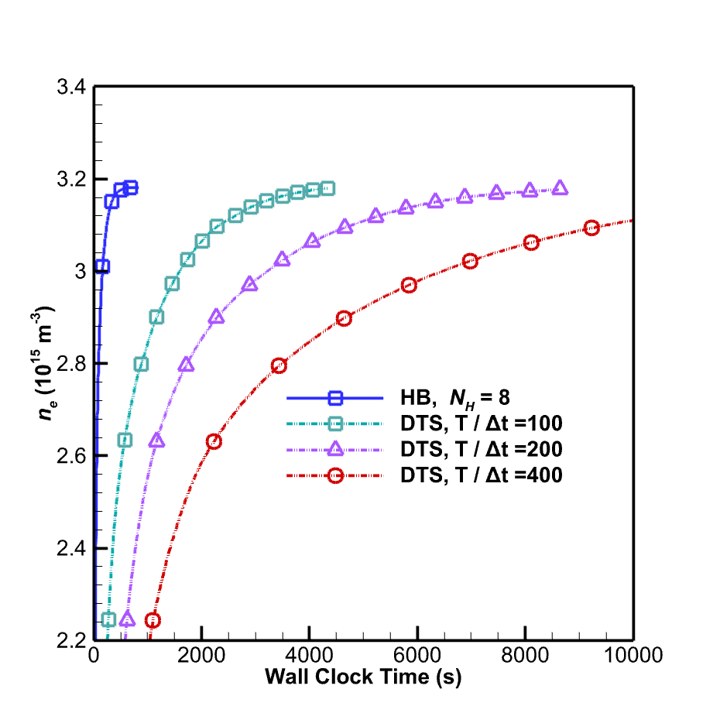}
    \caption{Convergence histories of the spatially averaged electron density: a wall-clock time comparison between the time-domain DTS method and the proposed HB method.}
    \label{fig:convergence_comparison}
\end{figure}

Beyond physical fidelity, another advantage of the proposed time-domain HB method lies in its exceptional computational efficiency. Figure~\ref{fig:convergence_comparison} compares the convergence histories of the spatially averaged electron density against the physical wall-clock time for both the DTS and HB methods. 

The traditional DTS approach dictates resolving the macroscopic evolution over hundreds to thousands of RF cycles before periodic steady-state conditions are met. As depicted, the computational expense of the DTS method scales almost linearly with the temporal resolution: achieving full convergence requires approximately 4344 s, 8643 s, and 17793 s for $T/\Delta t = 100$, 200, and 400, respectively. In stark contrast, the time-domain HB method directly targets the coupled steady-state solution across all temporal collocation points via pseudo-time relaxation. 

Executed on a single core of the aforementioned processor, the $N_H=8$ HB configuration reaches deep convergence in merely $842.5$~s. When evaluated against the rigorously verified DTS baseline ($T/\Delta t=200$), the HB approach delivers an impressive order-of-magnitude acceleration with a $10.26$-fold speedup. Even when compared to the coarsest DTS setup ($T/\Delta t=100$), the HB method remains $5.16$ times faster. Such a dramatic reduction in computational overhead is achieved on a purely sequential architecture without compromising physical accuracy. This unequivocally highlights the superior algorithmic efficiency and practical utility of the proposed HB framework for RF CCP simulations.

\section{Conclusion}
\label{sec:conclusion}

In this study, a  time-domain harmonic balance (HB) framework was developed to accelerate the fluid simulation of RF CCPs. This work represents the first successful extension of the time-domain HB method to a fully coupled drift-diffusion-Poisson system incorporating the complete electron-energy balance. To overcome the severe numerical stiffness introduced by this complex energy coupling and the dense time-spectral operator, a robust implicit pseudo-time relaxation method was proposed based on a spatiotemporal operator-splitting strategy. By strictly confining the dense temporal matrix inversion to the local cell level and incorporating a temporally decoupled semi-implicit Poisson update, the solver effectively decoupled the highly nonlinear governing system. This approach avoided the assembly of massive global Jacobian matrices while ensuring strong numerical stability against both chemical stiffness and the strict dielectric relaxation limit.

The proposed framework was rigorously validated against a standard one-dimensional argon CCP benchmark. Spectral analysis and convergence tests demonstrated that retaining eight harmonics ($N_H=8$) is perfectly sufficient to resolve the periodic RF discharge. Evaluated across all discrete temporal collocation points, the HB solution exhibited excellent agreement with the conventional DTS reference, successfully capturing the quasi-steady bulk plasma and the highly nonlinear transient sheath dynamics. The high fidelity of the proposed formulation naturally extended from the charged-particle continuity equations to the full electron-energy balance, with macroscopic relative errors remaining strictly below $0.3\%$.

Beyond physical accuracy, the time-domain HB method demonstrated exceptional computational efficiency. By directly solving for the coupled steady-state variables and completely bypassing the exhaustive physical transients, the proposed solver achieved a greater than 10-fold acceleration compared to the fully converged DTS baseline, and remained over 5 times faster even when compared to the coarsest time-marching setup. Crucially, this dramatic reduction in computational overhead was realized using a purely sequential single-core execution. In summary, this work confirms that the time-domain HB framework provides a high-fidelity, memory-efficient, and computationally superior alternative to conventional time-marching methods, showing tremendous potential for the routine parameter optimization and iterative design of RF plasma reactors.

\section*{Acknowledgments}
The current research is supported by National Science Foundation of China (92371107) and Hong Kong research grant council (16208324).

\bibliographystyle{main}
\bibliography{main}

\end{document}